\def\kms{\relax \ifmmode {\,\rm km\,s}^{-1}\else \,km\,s$^{-1}$\fi}    
   
\def\farcs{\hbox{$.\!\!^{\prime\prime}$}}   
   
\def\arcmin{\hbox{$^\prime$}}   
\def\arcsec{\hbox{$^{\prime\prime}$}}   
\def\secd#1.#2{ #1\farcs#2 }               

\def\mincir{\ \raise-2.truept\hbox{\rlap{\hbox{$\sim$}}\raise5.truept   
    \hbox{$<$}\ }}   
\def\magcir{\ \raise-2.truept\hbox{\rlap{\hbox{$\sim$}}\raise5.truept   
    \hbox{$>$}\ }}   
\def\gr{$^\circ$}

\def\nii{[N~{\sc ii}]} 
\def\ni{[N~{\sc i}]} 
\def\neiii{[Ne~{\sc iii}]}   
\def\oiii{[O~{\sc iii}]}   
\def\oi{[O~{\sc i}]} 
\def\oii{[O~{\sc ii}]}   
\def\ha{H$\alpha$}  
\def\hb{H$\beta$}  
\def\hg{H$\gamma$}  
\def\hd{H$\delta$}  
    
\def\ariii{[Ar~{\sc iii]}}   
\def\ariv{[Ar~{\sc iv]}}   
\def\hii{H~{\sc ii}} 
\def\hi{H~{\sc i}} \def\heii{He~{\sc ii}} 
 
\def\hei{He~{\sc i}}  
\def\sii{[S~{\sc ii}]}   
\def\siii{[S~{\sc iii}]}
\def\cbeta{$c_{\beta}$} 
 
\def\sexa{Sextans~A} 
\def\sexb{Sextans~B} 
 
\documentclass[twocolumn]{aa}   
\usepackage{graphicx}        
\begin{document}   
\title{The chemistry of planetary nebulae and HII regions in the dwarf
galaxies Sextans A and B from deep VLT spectra\thanks{Based on
observations obtained at the 8.2~m VLT telescope (Paranal, Chile)
operated by ESO (Proposal 072.B-0395) and at the 4.2~m WHT telescope
(La Palma, Spain) operated by the Isaac Newton Group in the Spanish
Observatorio del Roque de Los Muchachos of the Instituto de
Astrofisica de Canarias.}}
  
\author{  
Laura Magrini \inst{1}, Pierre Leisy\inst{2,3}, Romano L.M. Corradi 
\inst{3}, \\ Mario Perinotto \inst{1}, Antonio Mampaso \inst{2}, 
Jos\'{e} M. V\'{\i}lchez\inst{4} } 
\offprints{L. Magrini\\    
e-mail: laura@arcetri.astro.it}   
\institute{  
Dipartimento di Astronomia e Scienza dello Spazio, Universit\'a di  
Firenze, L.go E. Fermi 2, 50125 Firenze, Italy  
\and   
Instituto de Astrof\'{\i}sica de Canarias, c. V\'{\i}a L\'actea s/n,  
38200, La Laguna, Tenerife, Canarias, Spain  
\and    
Isaac Newton Group of Telescopes, Apartado de Correos 321, 38700 Santa    
Cruz de La Palma, Canarias, Spain 
\and 
Instituto de Astrof\'{\i}sica de Andaluc\'{\i}a  (CSIC)
Apartado de Correos 3004, 18080 Granada,  Spain
}  
  
\date{Received date /Accepted date }   
  
\abstract{ 
{\bf Spectroscopic observations obtained with the VLT of one planetary nebula (PN)  in \sexa\ 
and of five PNe in \sexb\ and of several \hii\ regions in these two  
dwarf irregular galaxies are presented.}  The extended spectral coverage, from 320.0 to
1000.0~nm, and the large telescope aperture allowed us to detect a
number of emission lines, covering more than one ionization stage for
several elements (He, O, S, Ar).
The electron temperature diagnostic
\oiii\ line at 436.3~nm was measured in all six PNe and in
several \hii\ regions allowing for an accurate determination of the ionic
and total chemical abundances by means of the Ionization
Correction Factors method.  For the time being, these PNe are the
farthest ones where such a direct measurement of the electron
temperature is obtained.  In addition, all PNe and
\hii\ regions were also modelled using the photoionization code CLOUDY 
(Ferland et al. \cite{ferland}).  The physico-chemical properties of
PNe and \hii\ regions are presented and discussed.  A small dispersion
in the oxygen abundance of \hii\ regions was found in both galaxies:
12 + $\log$(O/H)=7.6$\pm$0.2 in \sexa, and 7.8$\pm$0.2 in \sexb.  For
the five PNe of \sexb, we find that 12 + $\log$(O/H)=8.0$\pm$0.3, with
a mean abundance consistent with that of \hii\ regions.  The only PN
known in \sexa\ appears to have been produced by a quite massive
progenitor, and has a significant nitrogen overabundance.  In
addition, its oxygen abundance is 0.4~dex larger than the mean
abundance of \hii\ regions, possibly indicating an efficient third
dredge-up for massive, low-metallicity PN progenitors.  The metal
enrichment of both galaxies is analyzed using these new data.
 } 
\authorrunning{Magrini et al.}    
\titlerunning{PNe and H~II regions in Sextans~A and B }     
\maketitle    
   
\begin{figure*} 
\resizebox{\hsize}{!}{\includegraphics[angle=270]{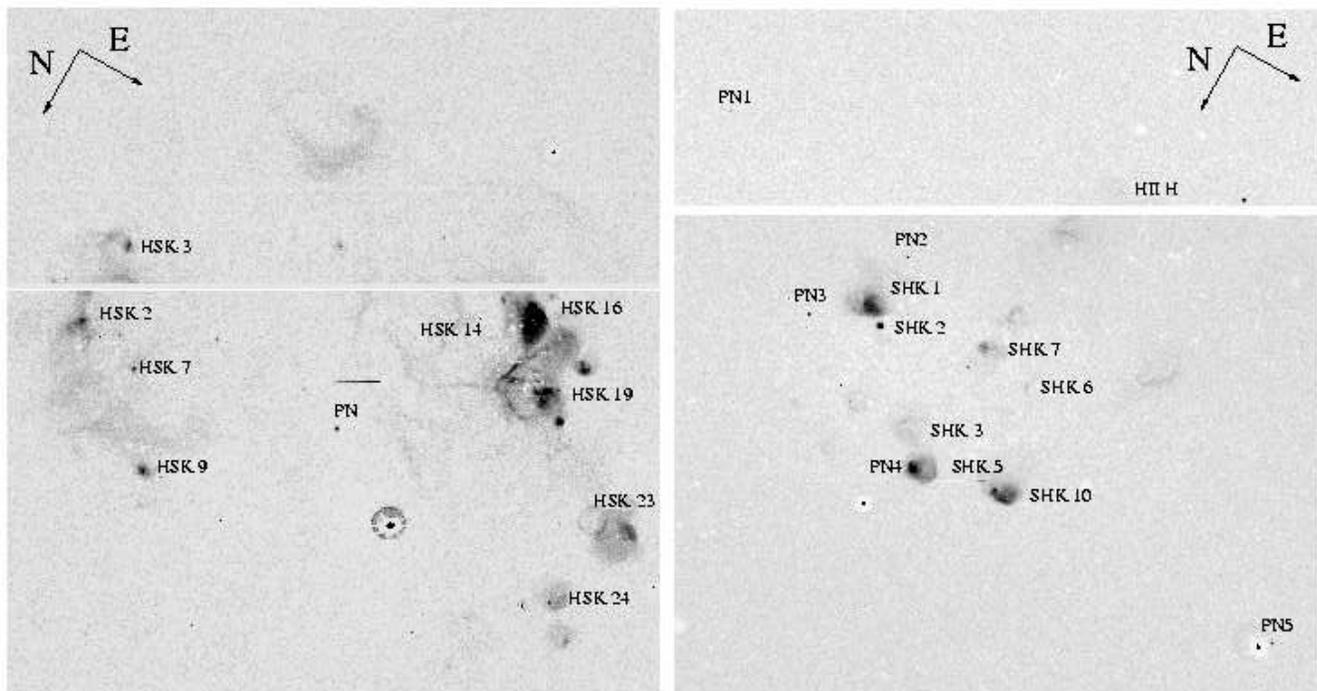}}  
\caption{VLT \ha\ - R images of \sexa\ (left) and \sexb\   
(right). The f.o.v. is 2\arcmin$\times$2\arcmin. The \hii\ regions and
PNe observed spectroscopically are indicated with their identification
numbers as in Tab.~\ref{Tab_pos_hii} for \hii\ regions and as in Magrini
et al. (\cite{m02}, \cite{m03}) for PNe.} 
\label{fig_sexab} 
\end{figure*} 
 
\section{Introduction} 
 
\sexa\ and \sexb\ are both dwarf irregular galaxies (Ir V and Ir 
IV-V morphological types, respectively, cf. van den Bergh 
\cite{bergh}, hereafter vdB00) with approximately the same V luminosity  
and located at a very similar distance (1.3 Mpc, Dolphin et
al. \cite{dolphin} for \sexa\ and vdB00 for \sexb).  Their separation
in the sky is also relatively small ($\sim$ 10 degrees), which
corresponds to about 280~kpc at the adopted distance.  Moreover their
velocity difference is only 23 $\pm$6 km s$^{-1}$.  All these
properties suggest a common formation for these two galaxies, probably
together with NGC~3109 and the Antlia galaxy, also at a similar
distance and location in the sky.  Considering the mean distance of
the four galaxies from the barycentre of the Local Group (LG),
1.7~Mpc, this sub-group is located beyond the zero velocity surface of
the LG (cf. vdB00) and can be considered the nearest external group of
galaxies.  Then these galaxies are particularly interesting as they
represent a group of dwarf galaxies relatively isolated from giant
galaxies.
 
In dIr galaxies, star formation is generally active, as shown by the 
conspicuous number of \hii\ regions that they contain.  There are 
several photometric and spectroscopic studies of 
\hii\ regions in both galaxies.  Photometry of \hii\ regions in \sexa\ 
was obtained by Hodge (\cite{hodge74}), Aparicio \&
Rodr\'{\i}guez-Ulloa (\cite{aparicio}), and Hodge et
al. (\cite{hodge94}), the latter one being the most complete survey
with 25 \hii\ regions detected.  Hunter et al. (\cite{hunter93})
studied large ionized gas structures outside normal \hii\ regions,
like shells and filaments generated by the action of massive stars
through winds and supernova explosions.
Spectra of four \hii\ regions were obtained by Skillman et al. 
(\cite{skillman89}), from which an oxygen abundance was derived.  The 
brightest \hii\ regions of \sexb\ were first catalogued by Hodge 
(\cite{hodge74}).  Strobel et al. (\cite{strobel}) classified twelve 
\hii\ regions, whereas oxygen abundance was derived for four \hii\ 
regions by Skillman et al. (\cite{skillman89}). Note, however, that 
these spectroscopic studies were generally not deep enough (except for 
one or two objects) to allow direct measurement of the nebular 
electron temperature via detection of the \oiii\ 436.3~nm line, 
causing an important source of uncertainty in the determination of
their chemical abundances. 

Planetary nebulae (PNe) are known to be good tracers of
intermediate-age stellar populations in galaxies.  One PN was
discovered in \sexa\ by Jacoby \& Lesser (\cite{jacoby81}), and then
reidentified by Magrini et al. (\cite{m03}).  In \sexb, five PNe are
known (Magrini et al. \cite{m02}).  To date, no spectroscopic studies
were done of these PNe.
 
As remarked by Mateo (\cite{mateo}) ``No two LG dwarfs have the same
star-formation history''. This is true not only when comparing
galaxies of different morphological type, luminosity, or mass, but
also when comparing galaxies which are very similar in many aspects,
as these two dwarf irregulars.  In fact, in spite of a probable common
origin, \sexa\ and \sexb\ show a different star formation history as
indicated by the different amounts of stars in the various
evolutionary phases, and also reflected in the different number of PNe
and \hii\ regions observed. \sexa\ possesses a conspicuous old stellar
population (10-14~Gyrs ago) and shows evidence for recent star
formation starting about 1-2~Gyrs ago, whereas the amount of star
formation at intermediate ages (2-10~Gyrs ago) was modest (see also
Dolphin et al. \cite{dolphin}).  On the other hand, \sexb\ exhibits a
very strong star formation at recent and intermediate ages (1-4~Gyrs
ago), together with a very old population (cf. Mateo \cite{mateo}).
The different numbers of PNe, 5 in \sexb\ vs. 1 in \sexa, was
interpreted by Magrini et al. (\cite{m02}) to be consistent with the
different star formation histories of the two galaxies for the age
range covered by the PN progenitors, \sexb\ showing a stronger star
formation rate than \sexa\ during the past 2-10~Gyrs.  The different
star formation history might imply a different chemistry.  In this
paper, we analyze the chemical content of \sexa\ and \sexb, presenting
spectroscopic data of PNe, whose progenitors belong to old and
intermediate age populations, and \hii\ regions, representative of the
youngest population.
 
The paper is organized as follows.  Sect.\ref{sect_obs} describes the
observations and the data reduction.  Sect.\ref{sect_chem} presents
the chemical abundances obtained for PNe and \hii\ regions, and
discusses the implications on the properties and evolution of the
galaxies.  In Sect.\ref{sect_pna} the notable case of the PN in
\sexa\ is examined. The summary is in Sect.\ref{sect_summ}. 
 
\begin{table} 
\caption{ 
The observed \hii\ regions in \sexa\ and \sexb.  The centres of
the slits are reported.  The adopted names of \hii\ regions are from
Hodge at al. (\cite{hodge94}) for Sextans A and from Strobel et
al. (\cite{strobel}) for Sextans B. The HII region not identified in
previous works is labeled as \hii~H. }
\begin{center} 
\renewcommand{\arraystretch}{1.4} 
\setlength\tabcolsep{5pt} 
\begin{tabular}{lll} 
\hline\noalign{\smallskip} 
Name   & RA & Dec\\ 
        &\multicolumn{2}{c}{J2000.0}     	 	\\	 
\noalign{\smallskip} 
\hline\hline 
\noalign{\smallskip} 
Sextans A&&\\
HKS2   & 10:10:53.6    & -4:41:16.9 	\\
HKS3    & 10:10:53.8    & -4:41:57.2 	\\
HKS7    & 10:10:55.4    & -4:41:11.0 	\\
HKS9    & 10:10:57.0    & -4:40:30.6 	\\
HKS14   & 10:11:04.1    & -4:42:49.3 	\\
HKS16  & 10:11:05.6    & -4:42:31.5 	\\
HKS19  & 10:11:06.9    & -4:42:17.8 	\\
HKS23  & 10:11:09.2    & -4:40:53.2 	\\
HKS24  & 10:11:09.6    & -4:41:36.3 	\\
\noalign{\smallskip} 
\hline\hline 
\noalign{\smallskip}
Sextans B&&\\ 
SHK1 & 9:59:57.7        & 5:19:36.7 	\\
SHK2 & 9:59:58.3        & 5:19:46.3 	\\
SHK3 & 9:59:59.9        & 5:20:17.4 	\\
SHK5 & 10:00:00.4       & 5:20:25.2 	\\
SHK7 & 10:00:00.8       & 5:19:38.3 	\\
SHK6 & 10:00:02.2       & 5:19:44.8 	\\
SHK10& 10:00:03.1      & 5:20:31.8 	\\
\hii~H& 10:00:05.3     & 5:17:58.7 	\\
\noalign{\smallskip} 
\hline 
\noalign{\smallskip} 
\end{tabular} 
\end{center} 
\label{Tab_pos_hii} 
\end{table} 

\section{Observations and data reduction} 
\label{sect_obs} 

All PNe known in these galaxies (1 in \sexa\ and 5 in \sexb) and a
sample of \hii\ regions (9 and 8, respectively) have been observed in
December 2003 with the 8.2~m VLT (ESO, Paranal) equipped with the
FORS2 spectrograph in multi-object spectroscopy (MOS) observing mode.
Pre-imaging, needed to create the MOS masks, was first obtained
through an \ha\ filter (central wavelength 656.3~nm, FWHM 6.1~nm) and a
R filter (655/165.0~nm).  The exposure times for each galaxy were
2$\times$120~s in \ha\ and 2$\times$10~s in R.  In the
continuum-subtracted \ha$-$R frames, emission-line objects were
identified and their position used to build two MOS masks, one per
galaxy.  

Spectroscopy was then secured during four consecutive nights (22-25
December).  The seeing during the observations was 0$''$.8. The MOS
mask used to observe \sexa\ contained 32 slits and the one for
\sexb\ 28 slits.  The slits had a width of 0.8\arcsec, while their
length varied from 3\arcsec\ to 25\arcsec, according to the size of
the object to be observed and to the crowding of the area of the
galaxy.  The slits corresponding to compact objects like PNe were
extended enough to include also some adjacent sky spectrum.  Several
independent slits were used to obtain sky spectra to be subtracted
from those of extended \hii\ regions, which completely fill their
slitlets.  Spectra were obtained using FORS2 with the 300V and 300I
gratings, providing 'blue' and 'red' spectra with a total coverage
from $\sim$320.0~nm to $\sim$1000.0~nm at a reciprocal dispersion of
0.3 nm~pix$^{-1}$ and an effective resolution of 0.6~nm. For each
galaxy, a total of eight exposures were taken: three with the 300V
grating (total exposure time 5400~s), three using the 300I one
(3600~s), and also two short exposures for each grating (60~s) to
avoid saturation of the most intense emission lines, as \ha\ and
\oiii\ 500.7~nm. A spectrophotometric standard star, GD~108, was observed
once in blue and red during two consecutive nights. The data were
reduced using the IRAF LONGSLIT tasks and the MIDAS MOS and LONG
packages.

\subsection{Images} 
 
The \ha-R FORS2 images of \sexa\ and \sexb\ are shown in 
Fig.~\ref{fig_sexab}.  All PNe in Magrini et al. (\cite{m02}, 
\cite{m03}) were re-identified, together with a number of known HII 
regions. 
Compared to previous studies (Hodge et al. \cite{hodge94}, Strobel et
al. \cite{strobel}), several new unresolved \ha\ emission-line objects
were found in both galaxies. When comparing their coordinates with the
images in Magrini et al. (\cite{m02}, \cite{m03}), no associated
\oiii\ emission was found, and therefore we do not consider 
them as new candidate PNe.  We observed spectroscopically two of them,
one in \sexa\ and one in \sexb, confirming that they are stars with a
very strong \ha\ line in emission.  A detailed analysis of these
objects via colour-colour diagrams and spectroscopy will be presented
in a forthcoming paper.

In Fig.~\ref{fig_sexab}, all PNe and \hii\ regions observed
spectroscopically are marked with their identification numbers as in
Tab.~\ref{Tab_pos_hii} for \hii\ regions and as in Magrini et
al. (\cite{m02}, \cite{m03}) for PNe.

\begin{figure*} 
\resizebox{\hsize}{!}{\includegraphics[width=3 truecm, angle=270]{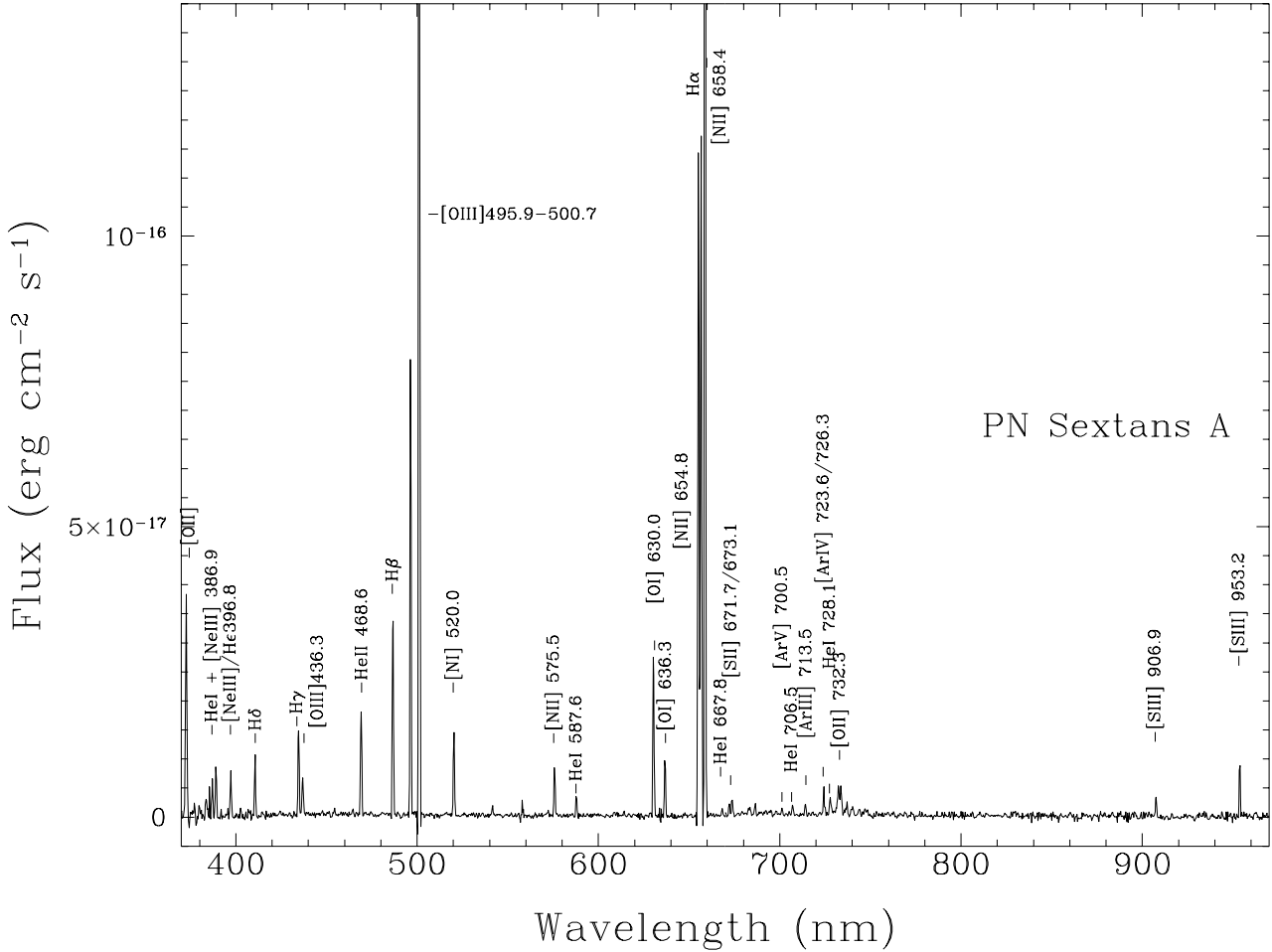}}  
\caption{VLT spectrum of the planetary nebula in \sexa.} 
\label{fig_s_pna} 
\end{figure*} 
 
\begin{figure*} 
\resizebox{\hsize}{!}{\includegraphics[width=6 truecm, angle=270]{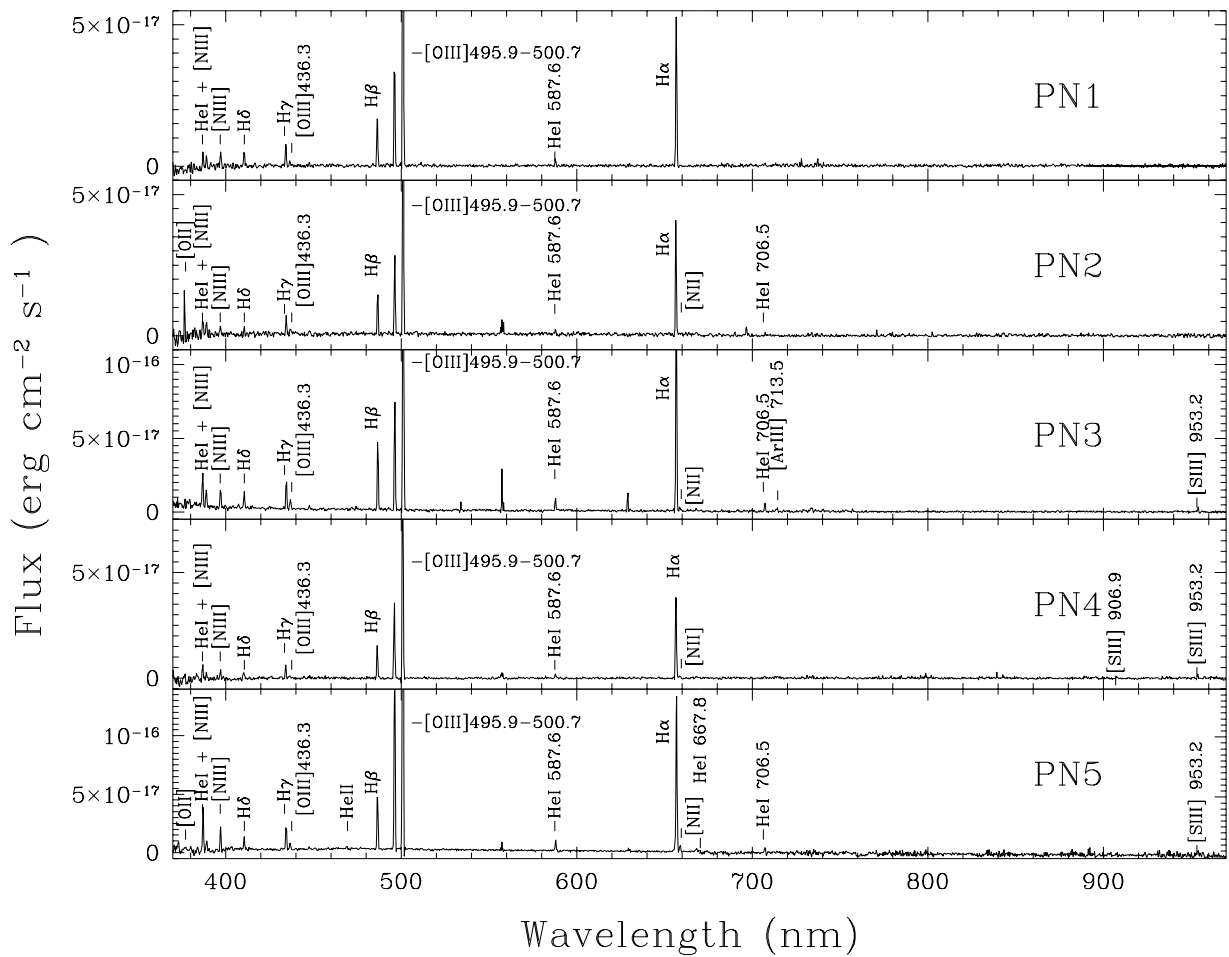}}  
\caption{VLT spectra of the five planetary nebulae in Sextans B.  
Identification numbers come from Magrini et al. (\cite{m02})} 
\label{fig_s_pnb}
\end{figure*} 
 
\begin{figure*} 
\includegraphics[width=14 truecm, angle=270]{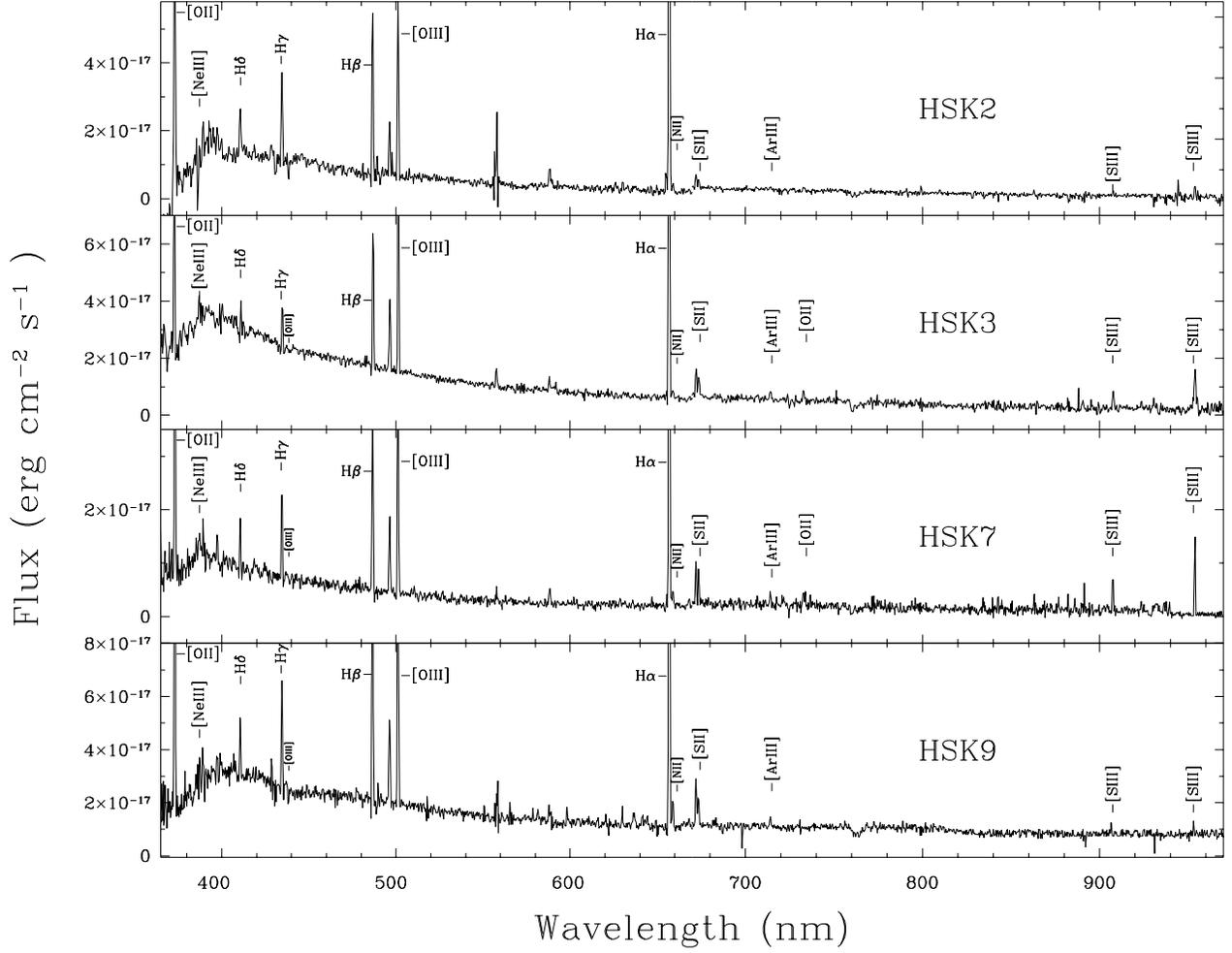}
\caption{VLT spectra of four \hii\ regions in \sexa.} 
\label{fig_s_hiia1} 
\end{figure*} 

\begin{figure*} 
\includegraphics[width=14 truecm, angle=270]{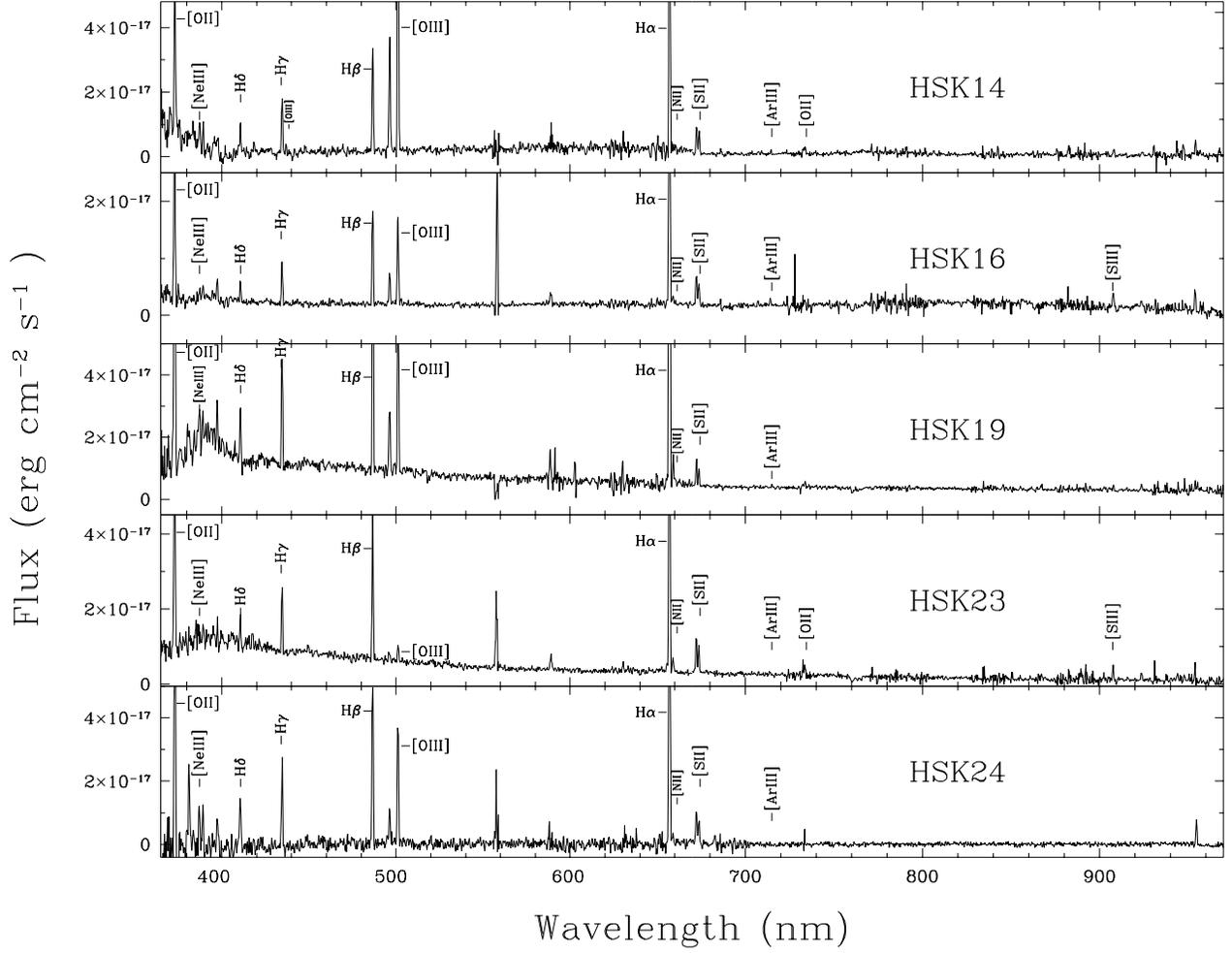}
\caption{VLT spectra of  another five \hii\ regions in \sexa.} 
\label{fig_s_hiia2} 
\end{figure*} 
 
\begin{figure*} 
\includegraphics[width=14 truecm, angle=270]{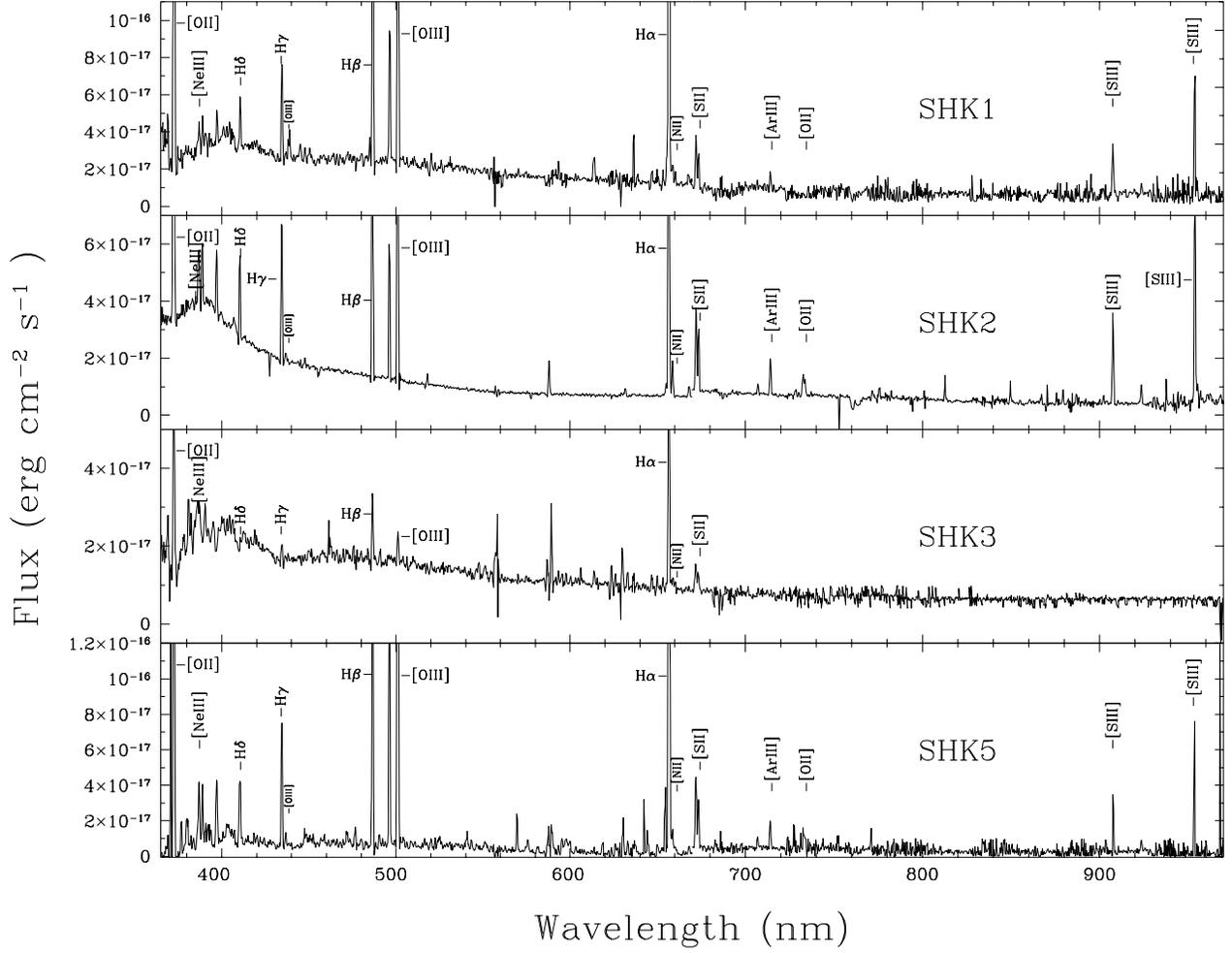} 
\caption{VLT spectra of four \hii\ regions  in \sexb.}  
\label{fig_s_hiib1} 
\end{figure*} 
 \begin{figure*} 
\includegraphics[width=14 truecm, angle=270]{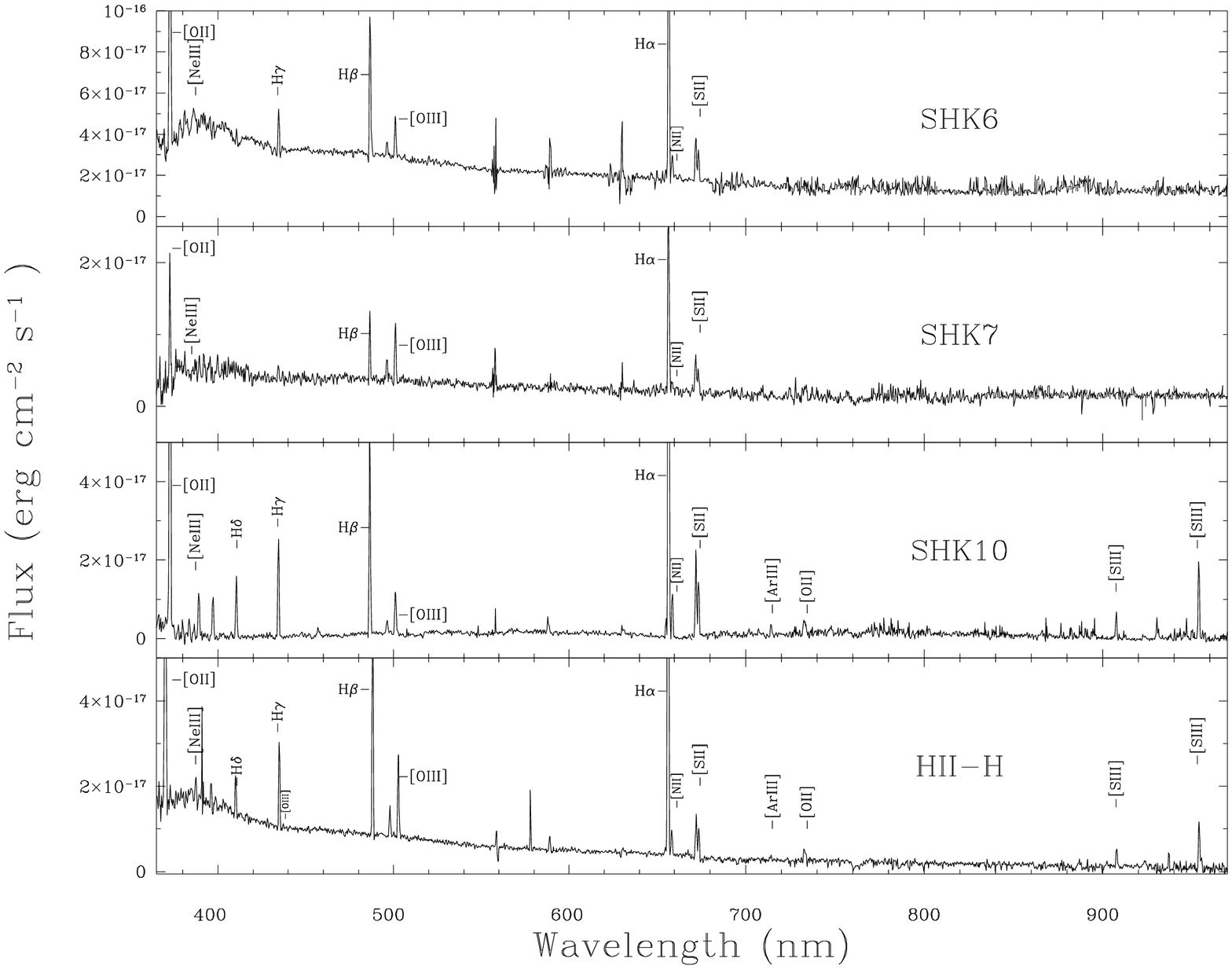} 
\caption{VLT spectra of another four \hii\ regions  in \sexb.}  
\label{fig_s_hiib2} 
\end{figure*} 
 
\subsection{Spectra} 
 
An important feature of this spectroscopic study is the homogeneous
set of observational data and analysis method used to measure chemical
abundances for both PNe and \hii\ regions.  Furthermore, the VLT
allows us to obtain spectra which are deep enough to measure most
basic diagnostic emission lines in spite of the significant distance
of the galaxies. The electron temperature diagnostic \oiii\ 436.3 was
measured in all PNe, and in seven \hii\ regions out of a sample of
17. In few other \hii\ regions (four in \sexa\ and one in \sexb), an
upper limit to the flux of the \oiii\ 436.3 line could be derived.
The broad spectral coverage, from 320.0 to 1000.0~nm, allowed us the
detection of many recombination (\hi, \hei, \heii) and forbidden lines
of several ions of O (\oi, \oii, \oiii), S (\sii,\siii), Ar (\ariii,
\ariv), Ne (\neiii), and N (\nii).  Note that the \siii\ lines at
906.9 and 953.2~nm are generally blended with saturated atmospheric
absorption features. In this case, the heliocentric velocities of both
galaxies, 324~kms\ for \sexa\ and 300~\kms\ for \sexb, and the
excellent sky transparency at these wavelength in Paranal, allowed us
to use safely these lines to determine the S$^{++}$ abundances.
 
The spectra of the PNe are presented in Figs. \ref{fig_s_pna} and
\ref{fig_s_pnb}, and those of \hii\ regions in
Figs. \ref{fig_s_hiia1}, \ref{fig_s_hiia2}, \ref{fig_s_hiib1}, and
\ref{fig_s_hiib2}. Identification of some emission lines is
labeled. Note the large number of emission lines detected with a good
signal to noise ratio, and the relatively strong blue continuum
emission visible in a number of \hii\ regions, which is likely
produced by the exciting stars.

It should be noted that although the observations were carried
out at relatively low airmasses, the slits were not aligned exactly to
the parallactic angle at the time of the observations: for \sexa\ the
mean difference between the parallactic angle and the position angle
of the slits was 45\gr\ and the mean airmass was 1.15, while for
\sexb\ they were 35\gr\ and 1.25, respectively.  This implies that,
given the relatively narrow slit, some amount of light is lost
especially in the blue end of the spectra due to differential
atmospheric refraction.  In order to estimate these light-losses for
point sources like PNe, we have calculated the displacement, as a
function of wavelength, of the centroid of an object from the midpoint
of the slit along the direction perpendicular to its length, using the
figures in Filippenko (\cite{filippenko82}). In doing so, we consider
that sources are well centred at H$\alpha$, which is the passband at
which acquisition images were taken to centre objects in their
slitlets for spectroscopy. We have then calculated the amount of light
lost (relative to that at \ha) because of these displacements from the
slit centre, assuming a Gaussian point-spread functions with a FWHM
which is the seeing of the observing night, and including seeing
worsening toward bluer wavelengths.

As expected, the differential light-losses are negligible in the red
side of the spectra, while become as large as 75\%\ for
\sexa, and 90\%\ for Sex~B, at the shortest wavelength that we
consider, which is that of the \oii~372.7 doublet.  At \hg, they
become as small as 20\%\ and 25\%, respectively.  We have applied
these correction factors to the fluxes of blue lines, and attached an
associated error of one third of the correction factor, which is
propagated in all quantities (ionic abundances, ionization correction
factor) derived using the blue lines.
Note that the correction for the interstellar reddening applied to
the spectra would by itself at least partially compensates for the
atmospheric refraction, by `forcing' the Balmer lines to have the
theoretical unreddened flux ratios (although with an estimated
extinction value \cbeta\ larger than the real one).  Thus for \sexa\
the difference between the observed fluxes and the refraction
corrected fluxes, both after the reddening correction, amounts to
50\%\ at the wavelength of the \oii\ 3727 doublet and negligible at
\hg\ for \sexa, and of 60\%\ and 10\%, respectively, for
\sexb.

In addition, it is important to note that these quite large
corrections at the bluest wavelengths do not affect significantly our
chemical analysis. In fact, the oxygen total and ionic abundances of
the PNe where the \oii\ 372.7 lines were measured change only by a
factor much smaller than 0.1~dex, with respect to the values
calculated using the observed fluxes.  Moreover, once the fluxes were
corrected for the effect of the atmospheric refraction, we analyzed
the differences of O$^+$/O derived from either the \oii~372.7 or 733.0
lines, in those PNe which have both doublets measured.  The
differences resulted to be less than 15\%, which implies in the worst
case, i.e. in the PN of \sexa, a difference in the total oxygen
abundance of only 0.05~dex.

For extended objects like \hii\ regions, the effect of atmospheric
dispersion is in principle more complex, because light is lost from
one side of the slit but is ``gained'' from the other side.  However,
the extension of the spectra along the slit length does not show
significant changes in surface brightness and excitation on scales of
$\sim$1.5~arcsec (after seeing convolution), which is the size of the
displacement due to the atmospheric dispersion.  Assuming that the HII
regions are also similarly homogeneous in the direction perpendicular
to the slit at these scales, then the effect of the atmospheric
dispersion is negligible at all wavelengths, and therefore no
corrections have been applied to the observed nebular spectra.  



{\bf Line fluxes and their errors are listed in Tables \ref{Tab_s_pn} and
\ref{Tab_s_hii}.  Errors include the contributions of   
background noise, sky subtraction, and flux calibration uncertainties.
In the case of PNe, errors include  the
correction for the atmospheric dispersion were applied; they range
from 40$\%$ at 370~nm to less than 1$\%$ at 500~nm.}

Finally, the observed line fluxes were corrected
for the effect of interstellar extinction, adopting Mathis
(\cite{mathis}) law with $R_V$=3.1. \cbeta, which is the
logarithmic difference between the observed and de-reddened \hb\
fluxes, was determined comparing the observed Balmer I(\ha)/I(\hb)
ratio with its theoretical value.  For the spectra with the higher
signal to noise ratio, we considered also the comparison between the
observed and theoretical Balmer I(\hb)/I(\hg) and I(\hg)/I(\hd)
ratios. In these cases \cbeta\ is the weighted average of the three
determinations.

 \subsection{Radial velocities} 

High-resolution spectroscopic data of three PNe of \sexb\ were
taken on April, 2002, with AF2/WYFFOS, a multi-object, wide field,
fibre spectrograph working at the prime focus of the 4.2~m WHT (La
Palma).  We used the Echelle grating with a central wavelength of
493.0~nm and a reciprocal dispersion of 0.028 nm/pixel.  The total
exposure time was 5400~s.  The reduction was done with the IRAF
DOFIBER package.
  
The high resolution WHT spectra allowed us to measure the Doppler
shift of the \oiii\ doublet lines, and in one case also of
\hb, in three PNe of \sexb, namely PN~1, PN~4 and PN~5. 
The heliocentric radial velocities are listed in Tab.\ref{Tab_s_pn}.
They are consistent, within the errors, with the heliocentric radial
velocity of \sexb\ (300 km/s from NED) and with the \hi\ observations
by Huchtmeier et al. (\cite{hucht}), confirming that the PNe belong
indeed to the galaxy.
 
\begin{table*} 
\caption{Observed line fluxes $F_\lambda$ of PNe  
in \sexa\ and \sexb, on a scale where F(\hb)=100.0, and after
correcting for atmospheric dispersion effects. `:' indicates uncertain
values, and `$<$' upper limits. Associated errors are described in the
text.  Some nebular quantities are also reported: \cbeta, the electron
density and temperatures T$_e$(\oiii) and T$_e$(\nii), and the
heliocentric radial velocity of three PNe where it was measured.
}
\begin{center} 
{\scriptsize 
\renewcommand{\arraystretch}{1.4} 
\setlength\tabcolsep{5pt} 
\begin{tabular}{lllllll} 
\hline\noalign{\smallskip} 
Ion 		 & SexA-PN & SexB-PN1 & SexB-PN2 & SexB-PN3 & SexB-PN4 & SexB-PN5 \\ 
\noalign{\smallskip} 
\hline 
\noalign{\smallskip} 
\cbeta\		        & 0.10$\pm$0.01  & 0.10$\pm$0.02     	&0.02$\pm$0.02 	&0.13$\pm$0.01 	&0.17$\pm$0.03 	    &0.22$\pm$0.01 	\\ 
\hline 
N$_e$(cm$^{-3}$)	& 2700$\pm$100   & -        		& -         	& -        	& -                 &- \\ 
\hline  
T$_e$(\oiii) (K)	& 22300$\pm$1000 & 12800$\pm$2000    	& 16000$\pm$3000& 20000$\pm$2000& 11300$\pm$3000    & 12800$\pm$1000  \\ 
T$_e$(\nii) (K)	        & 13400$\pm$500  & -        		& -         	& -        	& -                 & - \\ 
\noalign{\smallskip} 
\hline 
\noalign{\smallskip} 
V (km/s)		&-& 310$\pm$15 &- &-   &310$\pm$15 &310$\pm$15 \\ 
 
\noalign{\smallskip} 
\hline 
\noalign{\smallskip} 
 
372.7   \oii\     &  176.$\pm$70 &  18:     	&	-	   &-	    	    &-	       		& 21.$\pm$10	  \\  
383.5    H9       &  8.$\pm$4    &  -       	&	8.$\pm$6   & 11.$\pm$6	    &-	       		& -	  \\  
386.8   \neiii\   &  19.$\pm$6   &  38.$\pm$10  &	28.$\pm$7  & 55.$\pm$5	    &43.$\pm$7	       	& 93.$\pm$5	  \\   
388.9    \hei, H8 &  28.$\pm$5   &  16.$\pm$8   &	21.$\pm$6  & 32.$\pm$5	    &24.$\pm$5	       	& 16.$\pm$4	  \\   
396.7    \neiii, H7& 27.$\pm$3   &  29.$\pm$7   &	22.$\pm$7  & 29.$\pm$4	    &37.$\pm$5	       	& 46.$\pm$3	  \\  
410.1    \hd\     &  31.$\pm$2   &  25.$\pm$7   &	19.$\pm$6  & 26.$\pm$4	    &26.$\pm$7	       	& 23.$\pm$3	  \\  
434.0    \hg\     &  48.$\pm$2   &  48.$\pm$4   &	45.$\pm$4  & 51.$\pm$3	    &51.$\pm$	       	& 45.$\pm$1.6	  \\  
436.3    \oiii\   &  23.$\pm$2   &  8.$\pm$4    &	13.$\pm$4  & 16.$\pm$4	    &7.$\pm$5		& 11.$\pm$1.9	  \\  
447.1    \hei     &  1.3$\pm$1   &  5.$\pm$4    &	9.$\pm$3   & 4.$\pm$2	    &-	       		& 5.$\pm$2	  \\  
468.6    \heii    &  51.$\pm$2   &  $<$1    	&	$<$3       & $<$2	    &$<$1	       	& 6.$\pm$2	  \\  
471.2	\hei,\ariv&-	    	 & -	        &	-	   & 2.$\pm$1	    &-	       		& -	  \\ 
474.0	\ariv	 & -	    	 & -	        &	-	   & 3.$\pm$2	    &-	       		& -	  \\ 
486.1    \hb\     &  100.$\pm$1  &  100.$\pm$4  & 	100.$\pm$4 & 100.$\pm$1.5   &100.$\pm$5	       	& 100.$\pm$2	  \\  
495.9    \oiii\   &  206.$\pm$1  &  202.$\pm$3  &	190.$\pm$4 & 182.$\pm$1.6   &256.$\pm$3	       	& 316.$\pm$1.8	  \\  
500.7	\oiii\	 &  611.$\pm$1   &  590.$\pm$2  &	570.$\pm$3 & 474.$\pm$1     &798.$\pm$3	       	& 940.$\pm$1.5	  \\  
520.0	\ni\	 &  38.$\pm$1    & -	       	&	-	   &-	    	    &-	       		& -	  \\  
541.1	\heii\	 &  4.2$\pm$1    & -	       	&	-	   &-	    	    &-	       		& -	  \\  
575.5	\nii\	 &  24.$\pm$1    & -        	&	-	   &-	            &-       		& 	  \\  
587.6	\hei\    &  8.$\pm$1     & 16.$\pm$2    &	17.$\pm$5  &15.$\pm$1	    &15.$\pm$3	       	& 22.$\pm$2	  \\  
630.0	\oi\     &  71.$\pm$1    & 3.$\pm$2     &	3.$\pm$1   &-	    	    &-	       		& -	  \\  
636.3	\oi\     &  24.$\pm$1    & -	        &	-	   &-	            &-	       		&-	  \\  
654.8	\nii\    &  291.$\pm$1.  & -	        &	-	   &-	            &-	       		&-	  \\  
656.3	\ha\     &  306.$\pm$1.5 & 302.$\pm$2   &	290.$\pm$3 &312.$\pm$1.3    &321.$\pm$3	       	& 332.$\pm$1.3	  \\    
658.4	\nii\	 &  835.$\pm$1.8 & 2:       	&	7:	   &6.$\pm$1.8	    &4.$\pm$3 	       	& 12.$\pm$1.2	  \\    
667.8	\hei\    &  3.$\pm$1.6   & 3.$\pm$2     &	3.$\pm$1.5 &4.$\pm$1	    &-	       		& 3.$\pm$1.5	  \\       
671.7	\sii\	 &  5.$\pm$1.6   &-	        &	-	   &-	    	    &-	       		&-	  \\       
673.1	\sii\	 &  8.$\pm$1.5   &-	        &	-	   &-	    	    &-	       		&-	  \\       
706.5	\hei\	 &  4.$\pm$1.6   & 6.$\pm$3     &	-	   &10.$\pm$4	    &4.$\pm$2	        &3.$\pm$1.5	  \\       
713.5	\ariii\	 &  3.$\pm$3     &-	        &	-	   &3.$\pm$2	    &-	       		&-	  \\       
728.1	\hei\    &  2.$\pm$2     &-	        &	-	   &-	    	    &-	       		&-	  \\       
732.5    \oii\    & 28.$\pm$2    & 5.$\pm$3     &	7.$\pm$3   &-	    	    &11.$\pm$2	       	&-	  \\      
906.9    \siii\   &  7.$\pm$3    &-	        &	-	   &-	    	    &2.$\pm$2	       	&-	  \\      
953.2	\siii\   &  19.$\pm$2    &-	        &	-	   &5:	    	    &5:		       	&17:	  \\    
\noalign{\smallskip} 
\hline 
\noalign{\smallskip}     
\end{tabular} 
} 
\end{center} 
\label{Tab_s_pn} 
 
\end{table*} 
 
\begin{table*} 
\caption{Observed line fluxes $F_\lambda$,  on a scale where
F(\hb)=100.0, and nebular parameters of the \hii\ regions in Sextans A
and Sextans B. Symbols are as in Tab.~1.}
\begin{center} 
{\tiny 
\renewcommand{\arraystretch}{1.4} 
\setlength\tabcolsep{5pt} 
\begin{tabular}{llllllllll} 
\hline\noalign{\smallskip} 
	  SexA		 & HSK2  & HSK3  & HSK7  & HSK9   &  HSK14 & HSK16  & HSK19    & HSK23  	& HSK24         	\\ 
\hline 			           	                                                                 			  	   
\noalign{\smallskip} 	           	                                                                 			  	   
\cbeta\		 	 & 0.0$\pm$0.01 & 0.23$\pm$0.01 &0.13$\pm$0.01 &0.0$\pm$0.01 & 0.05$\pm$0.02 &0.14$\pm$0.01 &0.05$\pm$0.01  & 0.21$\pm$0.01   	&0.0$\pm$0.02     	              \\ 
\hline 			           	                                                                 			  	   
N$_e$(cm$^{-3}$)	 & 100$\pm$100& 200$\pm$100&$<$100  & 400$\pm$200& $<$100  &$<$100   & $<$100    & $<$100	& 100$\pm$100         	\\ 
\hline 			           	                                                                 			  	   
T$_e$(\oiii) (K)	 & $<$12000&13000$\pm$3000    & 17000$\pm$4000& 14000:    & 14000$\pm$1000   &$<$14000 &$<$13000   & -     	& $<$13000 \\ 
\noalign{\smallskip} 
\hline 
\noalign{\smallskip} 
Ion   			 &  &  &  &  &  &  &  &  & \\
\noalign{\smallskip} 	           	                                                                 			  	   
372.7   \oii\     	 &240.$\pm$10   &  126.$\pm$10  & 205.$\pm$10 & 145.$\pm$9    & 145.$\pm$16  &213.$\pm$10   &179.$\pm$8       &  194.$\pm$8	&	160.$\pm$10     \\  
386.8   \neiii\   	 &11.$\pm$5     &  15.$\pm$8    & 8.$\pm$6    & 10.$\pm$6     & 28.$\pm$8    &8.$\pm$5      &15.$\pm$5        &  5.$\pm$3 	&	22.$\pm$8	\\   
388.9    \hei, H8 	 & -       	&  -       	& 9.$\pm$5    & 15.$\pm$5     & 21.$\pm$7    &15.$\pm$5     &8.$\pm$4         &  10.$\pm$3  	&	18.$\pm$7       \\   
396.7    \neiii, H7	 & -       	&  -       	& 14.$\pm$3   & 10:           &-             &20.$\pm$4     &17.$\pm$3        &  9.$\pm$2   	&	20.$\pm$8      \\  
410.1    \hd\     	 & 24.$\pm$3    &  20.$\pm$5    & 25.$\pm$3   & 22.$\pm$3     & 26.$\pm$3    &21.$\pm$2     &25.$\pm$2        &  25.$\pm$2  	&	28.$\pm$8	\\  
434.0    \hg\     	 & 42.$\pm$2    &  35.$\pm$2    & 46.$\pm$3   & 41.$\pm$3     & 47.$\pm$3    &46. $\pm$2    &42.$\pm$2        &  43.$\pm$2  	&	47.$\pm$5           \\  
436.3    \oiii\   	 & $<$1    	&  2$\pm$2     	& 3$\pm$3     & 2.$\pm$3      & 6.$\pm$3     &$<$1          &$<$1             &  -    		&	$<$1	       	\\  
468.6    \heii    	 & $<$1    	&  $<$1    	& $<$1        & $<$1          & 3:           &6:            &$<$1             &  $<$1 	&		$<$1	            \\  
486.1    \hb\     	 &100.$\pm$2    &  100.$\pm$2   & 100.$\pm$2  & 100.$\pm$2    & 100.$\pm$3   &100.$\pm$2    &100.$\pm$1       &  100.$\pm$2 	&   	100.$\pm$3       \\  
495.9    \oiii\   	 &27.$\pm$2     &  57.$\pm$2    & 39.$\pm$2   & 31.$\pm$2     & 115.$\pm$3   &22.$\pm$2     &24.$\pm$1        &  5.$\pm$3   	&	25.$\pm$3           \\  
500.7	\oiii\	 	 &84.$\pm$2     &  159.$\pm$2   & 120.$\pm$2  & 94.$\pm$2     & 340.$\pm$3   &65.$\pm$2     &72.$\pm$1.5      &  15.$\pm$2  	&	75.$\pm$3      	\\  
587.6	\hei\    	 & -       	&  7.$\pm$2    	& 10.$\pm$2   & 4.$\pm$2            & -            &-             &-                &  -    		&	-  \\  
630.0	\oi      	 & -       	&  -       	& -           & -             & 8.$\pm$2     &-             &-                &  -    		&	-    \\    
656.3	\ha\     	 &281.$\pm$2    &  335.$\pm$2   &312.$\pm$2   &281.$\pm$2     & 295.$\pm$3   &314.$\pm$2    &295.$\pm$3       &  329.$\pm$2 	&	281.$\pm$3     \\    
658.4	\nii\	 	 &7.$\pm$2      &  9.$\pm$3     &9.$\pm$2     &11.$\pm$2      & 7.$\pm$3     &8.$\pm$2      &10.$\pm$3        &  13.$\pm$2  	&	10.$\pm$3     \\    
671.7	\sii\	 	 &18.$\pm$2     &  16.$\pm$3    &17.$\pm$2    &17.$\pm$2      & 17.$\pm$3    &24.$\pm$2     &26.$\pm$2        & 22.$\pm$2   	&	26.$\pm$3	\\     
673.1	\sii\	 	 &14.$\pm$2     &  13.$\pm$3    &11.$\pm$2    &15.$\pm$2      & 11.$\pm$3    &16.$\pm$2     &16.$\pm$2        & 14.$\pm$2  	&	20.$\pm$3 \\       
713.5	\ariii\	 	 &4.$\pm$2      &  5.$\pm$3     &5.$\pm$2     &3.$\pm$2       & 3.$\pm$3     &6.$\pm$2      &2.$\pm$2         & 2.$\pm$2        &	1.$\pm$3 \\     
732.5    \oii\    	 &-        	&  13.$\pm$3    &8.$\pm$2     &-              & 14.$\pm$3    &-             &-                & 8.$\pm$2    	&	- \\     
906.9    \siii\   	 & 5.$\pm$2     &  18.$\pm$3    &16.$\pm$3    &7.$\pm$5       & 8.$\pm$2     &20.$\pm$3     &-                & 8.$\pm$2   	&	- \\     
953.2	\siii\   	 & 14.$\pm$2    &  50.$\pm$3    &42.$\pm$5    &18.$\pm$5      & -            &-             &-                & -     	&	-  \\    
\noalign{\smallskip} 
\hline 
\noalign{\smallskip}   
\hline 
  SexB		    	& SHK1    & SHK2      & SHK3    & SHK5     & SHK6   & SHK7              & SHK10 & \hii~H        &		\\ 
\noalign{\smallskip}                                                                      			                      
\hline 		                                                                           			                      
\noalign{\smallskip}                                                                      			                      
\cbeta\		    	& 0.0$\pm$0.02   &0.13$\pm$0.01        &0.30$\pm$0.05      &0.16$\pm$0.01        &0.30$\pm$0.03    &0.12$\pm$0.02                  &0.29$\pm$0.01    & 0.15$\pm$0.01                   &		\\ 
\hline 		                                                                                                          
N$_e$(cm$^{-3}$)	& $<$100  & 200$\pm$100& $<$100  & $<$100    & $<$100& 300$\pm$100        &$<$100 &$<$100           &		\\ 
\hline 		                                                                                                          
T$_e$(\oiii) (K)	& 13000:  & 15000:    & -       & 12000$\pm$2000     & -     & -                   & -     & $<$17000              & 		\\ 
\noalign{\smallskip} 
\hline 
\noalign{\smallskip} 
Ion   			 &  &  &  &  &  &  &  &  &  \\ 
\noalign{\smallskip}
372.7   \oii\       & 181.$\pm$7   & 350.$\pm$4     &390.$\pm$30   & 182.$\pm$8       & 210.$\pm$40   &324.$\pm$10        &413.$\pm$6	 &  292.$\pm$7       &   	\\ 
386.8   \neiii\     & 9.$\pm$3     & 29.$\pm$3      &-        	   & 17.$\pm$4        & -             &-                  &7.$\pm$3	 &  9.$\pm$3         &           \\ 
388.9    \hei, H8   & 8.$\pm$2     & 26.$\pm$2      &-        	   & 14.$\pm$3        & -             &-                  &23.$\pm$2	 & 11.$\pm$3         &           \\ 
396.7    \neiii, H7 & 10.$\pm$2    & 27.$\pm$1      &-        	   & 20.$\pm$2        & -             &-                  &20.$\pm$2	 & 12.$\pm$3         &           \\ 
410.1    \hd\       & 13.$\pm$2    & 34.$\pm$1      &-        	   & 22.$\pm$2        & -             &-                  &30.$\pm$2	 &  19.$\pm$2        &           \\ 
434.0    \hg\       & 27.$\pm$2    & 53.$\pm$1      &-        	   & 38.$\pm$2        & -             &-                  &47.$\pm$2	 &  40.$\pm$2        &           \\ 
436.3    \oiii\     & 2.$\pm$2     & 3.$\pm$2       &-        	   & 3.$\pm$2              & -             &-                  &-	   	 &  $<$1     &  		\\ 
468.6    \heii      &$<$1          &$<$1            &-        	   & $<$1             & -             &$<$1               &$<$1	 	 &  -          &   	\\  
486.1    \hb\       & 100.$\pm$3   & 100.$\pm$0.5   &100.$\pm$9    & 100.$\pm$2       & 100.$\pm$5    &100.$\pm$4         & 100.$\pm$1	 &  100.$\pm$1       &   	\\ 
495.9    \oiii\     & 50.$\pm$3    & 55.$\pm$0.5    &17.$\pm$7     & 86.$\pm$2        & 35.$\pm$5     &12.$\pm$4          &9.$\pm$1	 &  15.$\pm$1        &           \\ 
500.7	\oiii\	    & 147.$\pm$3   & 148.$\pm$0.5   &52.$\pm$7     & 264.$\pm$2       & 113.$\pm$5    &30.$\pm$4          &23.$\pm$1	 &   38. $\pm$1      &   	\\ 
587.6	\hei\       & -            & 16.$\pm$0.5    &-             & 5.$\pm$          & -             &-                  &-	 	 &   7.$\pm$1        &           \\ 
656.3	\ha\        & 272.$\pm$    & 312.$\pm$0.5   &351.$\pm$4    & 319.$\pm$1.5     &351.$\pm$8     &310.$\pm$2         &350.$\pm$0.5	 &  316.$\pm$1        &   	\\ 
658.4	\nii\	    & 5.$\pm$2    & 15.$\pm$0.5    &25.$\pm$4     & 8.$\pm$1         &28.$\pm$7      &23.$\pm$2          &22.$\pm$1	 &  10.$\pm$1        & 		\\ 
667.8   \hei\       & -       	   & 5..$\pm$0.5    &-        	   & -                & -             &-                  &- 	 	 &  -          &           \\ 
671.7	\sii\	    & 15.$\pm$2    & 27.$\pm$1      &59.$\pm$4     & 21.$\pm$2        &75.$\pm$8      &29.$\pm$3          &33.$\pm$1	 &  22.$\pm$1       &  		\\ 
673.1	\sii\	    & 9.$\pm$2     & 22.$\pm$1      &43.$\pm$4     & 15.$\pm$2        &53.$\pm$8      &24.$\pm$3          &21.$\pm$1	 &  16.$\pm$1       &         	\\ 
706.5   \hei\       & -            & 4.$\pm$1       &-        	   & -                & -             &-                  &- 	 	 &  -          &           \\ 
713.5	\ariii\	    & 4.$\pm$2     & 12.$\pm$1      &-        	   & 2.$\pm$2              &-              &6.$\pm$3           &5.$\pm$1	 &  2.$\pm$1          &   	\\ 
732.5    \oii\      & -       	   & 13.$\pm$1      &-        	   & 9.$\pm$2         &-              &-                  &12.$\pm$2	 &  9.$\pm$1         &           \\ 
906.9    \siii\     & 10.$\pm$3    & 16.$\pm$1      &-        	   & 18.$\pm$2        &-              &-                  &12.$\pm$2	 &  8.$\pm$2         &           \\ 
953.2	\siii\      & 24.$\pm$3    & 41.$\pm$1      &-        	   & 43.$\pm$2        &-              &-                  &36.$\pm$2     &  21.$\pm$2       &           \\ 
\noalign{\smallskip}	 
\hline 
\noalign{\smallskip}     
\end{tabular} 
} 
\end{center} 
\label{Tab_s_hii} 
 
\end{table*}

\section{Chemical abundances of planetary nebulae and \hii\ regions} 
\label{sect_chem} 
 
\subsection{ICF method} 
 
Chemical abundances were derived from the spectra by means of
Ionization Correction Factors (ICF) method, following Kingsburgh \&
Barlow (\cite{kb94}), with evaluation of the errors as in Corradi et
al. (\cite{corradi97}) and Perinotto \& Corradi (\cite{perinotto98}).
As mentioned above, the \oiii\ 436.3 emission line was always
measurable in all PNe with a sufficiently high signal to noise ratio,
thus their \oiii\ electron temperature, T$_{e{\rm[O~III]}}$, could be
determined.  These are listed in Tab.~\ref{Tab_s_pn}; their high
values are typical of low-metallicity PNe (c.f.  Stasinska
\cite{stasinska02}).  The \oiii\ 436.3 line was also measured in four
\hii\ regions of \sexa\ and in three \hii\ regions of \sexb, while its upper 
limit was estimated in four \hii\ regions of \sexa\ and one of \sexb.
Another temperature diagnostic line, \nii\ 575.5~nm line, was observed
in the PN of Sextans~A, and a corresponding electron temperature
T$_{e{\rm[N~II]}}$ was determined.  In this object, T$_{e{\rm[O~III]}}$ was
used in computing chemical abundances with the ICF method for ions
which are ionized twice or more times, and T$_{e{\rm[N~II]}}$ for
species only ionized once.  The large difference between
T$_{e{\rm[O~III]}}$ and T$_{e{\rm[N~II]}}$ is typical of high
excitation PNe and it was also found in several Galactic PNe
(i.e. McKenna et al. \cite{mckenna}).  Kingsburgh \& Barlow
(\cite{kb94}) examined how the ratio between T$_{e{\rm[O~III]}}$ and
T$_{e{\rm[N~II]}}$ behaves with nebular excitation, plotting this
ratio versus the \heii\ 4686 flux.  They found a linear relation
between these two quantities, as expressed in eq. 3 of their paper.
The PN of \sexa, which is a very high excitation object, follows this
relationship.
 
The \siii\ near-infrared emission lines were detected in 4 PNe and 12 
\hii\ regions.  These lines allow a better accuracy in the measurement 
of sulphur abundance, especially at low metallicities where \sii\ 
lines are extremely faint. 
 
The ICF method was applied to the whole sample of PNe and to 
\hii\ regions where T$_{e{\rm[O~III]}}$ was measured. 
Formal errors on the ICF abundances were computed taking into account
the uncertainties in the observed fluxes and in the electron
temperatures. 
{\bf A discussion of the error propagation computed with the ICF method 
is in Corradi et al. (\cite{corradi97})}.

\subsection{Modelling with CLOUDY} 
 
PNe and \hii\ regions were also modelled using the photoionization
code CLOUDY 94.00 (Ferland et al. \cite{ferland}), assuming a
blackbody central star with effective temperature derived using the
Ambartsumian's (\cite{ambar}) or Gurzadyan's (\cite{gurzadyan})
methods, and a spherical nebula with constant density.  Density was
derived from the \sii\ 671.7/673.1 flux ratio, which was measured in
all \hii\ regions and in the PN of \sexa. For PNe in \sexb, it was
assumed equal to 3000~cm$^{-3}$, which can be considered a typical
value for Galactic PNe (see e.g. Stanghellini \& Kaler
\cite{stanghe89}).  In addition, we made some tests varying the density of 
$\pm$1000~cm$^{-3}$ with both the ICF and CLOUDY methods, obtaining 
negligible changes in the computed abundances. 

In the model, we include dust typical of Galactic PNe and of the Orion 
nebula, for PNe and \hii\ regions, respectively.   
The accuracy of the CLOUDY method was tested in Magrini et 
al. (\cite{M04}), who applied it to seven bright Galactic PNe using 
only the emission lines typically present in extragalactic PN spectra, 
and compared the results with the chemical abundances obtained with 
the ICF method by KB94.  A good agreement was found in the helium 
and metals abundances.  For further details 
on the modelling procedure see Magrini et al. (\cite{M04}) and 
Perinotto et al. (\cite{peri04}). 
 
\subsection{Results} 
 
The chemical abundances derived for the current sample with the ICF
and CLOUDY methods are in general agreement with each other, showing
the following r.m.s. differences in both galaxies: $\pm$0.01 for He/H
(linear), $\pm$0.15~dex for N/H, $\pm$0.1~dex for O/H, $\pm$0.1~dex for Ne/H,
$\pm$0.1~dex for S/H,$\pm$ 0.1~dex for Ar/H for both PNe and \hii\ regions.
 
The total and ionic abundances, and the ICF of PNe and \hii\
regions are reported in Tab.
\ref{Tab_ion_pn} and \ref{Tab_ion_sexabhii}. 
The total abundances for the whole sample of objects computed
with both methods, and their mean values are shown in
Tab.~\ref{Tab_c_pn} for PNe and in Tab.~\ref{Tab_c_hii_a} and
\ref{Tab_c_hii_b} for \hii\ regions. \hii\ regions where only the
CLOUDY method was applied because of the absence of a reliable
determination of the electron temperature are marked with a ``$*$'' in
Tab.~\ref{Tab_c_hii_a} and \ref{Tab_c_hii_b}.  For \sexb\ PN~3, ICF(N)
and ICF(S) were not computed because the \oii\ lines needed to compute
them were not detected, and thus for these elements only the results
from CLOUDY are quoted.

In the case of \hii\ regions, the ICF of He does not take into account
the neutral helium, which might be a considerable fraction of the
total helium abundance. However, for the \hii\ regions where
T$_{e{\rm[O~III]}}$ was measured, and therefore where the abundances
with the ICF method were computed (HKS3, HKS7, HKS9, HKS14 of \sexa\
and SHK1, SHK2, SHK5 of \sexb), the comparison between the results of 
ICF and CLOUDY methods shows that the mean contribution of neutral
helium is smaller than 20\%.  This relatively small contribution makes
unnoticeable the difference between the two results within our present
errors.
  
According to the definition by Kingsburgh \& Barlow (\cite{kb94}) for
Galactic PNe, the PN in \sexa\ would be classified as a type I PNe in
the Peimbert scheme ($\log$(N/O)$>$-0.1), while all PNe in \sexb\
would be non-Type~I. However, it is clear that this distinction is
hardly applicable to extragalactic PNe, which span a much larger range
of metallicities.

\begin{table*} 
\caption{ Total and ionic chemical abundances, and ionization correction factors  of \sexa\ and \sexb\ PNe,  
computed with  the ICF method. } 
\begin{center} 
\renewcommand{\arraystretch}{1.4} 
\setlength\tabcolsep{5pt} 
\begin{tabular}{lllllll} 
\hline\noalign{\smallskip} 
					& SexA-PN 	  & SexB-PN1      & SexB-PN2    & SexB-PN3      & SexB-PN4      & SexB-PN5 \\ 
\noalign{\smallskip} 
\hline 
\noalign{\smallskip} 
He$^+$/H$^+$$\times$10$^{2}$		&3.6$\pm$1.1	  & 9.8$\pm$1.5	  & 9.3$\pm$1.5	& 7.4$\pm$1.5	&9.4$\pm$1.3	&12.8$\pm$2.		\\
He$^{++}$/H$^+$$\times$10$^{2}$		&4.8$\pm$0.5	  & 0.09$\pm$0.05 & 0.3$\pm$0.1	& 0.2$\pm$0.1	&0.1$\pm$0.1	&0.5$\pm$0.1		\\
\noalign{\smallskip} 
He/H$\times$10$^{2}$			&8.4$\pm$1.6	  & 9.9$\pm$1.5	  & 9.6$\pm$1.6	& 7.6$\pm$1.6	&0.95$\pm$1.4	&13.3$\pm$2.1		\\
\hline

\noalign{\smallskip} 
N$^0$/H$^+$$\times$10$^{5}$		&3.9$\pm$ 1.2 	  & -		  &-		 &-		&-		&-		\\
N$^+$/H$^+$$\times$10$^{5}$		&7.1$\pm$ 0.9	  & 0.008:	  &0.003:	 &0.03$\pm$0.01 &0.015$\pm$0.01 &0.04$\pm$0.02		\\
ICF(N)					&3.0	  	  & 72.	  	  &19.		 &-		&45.		&130.		\\
\noalign{\smallskip} 
N/H$\times$10$^{5}$			&22.$\pm$10.	  & 0.5:	  &0.5:		 &$>$0.03	&0.7$\pm$0.5	&5.$\pm$3.		\\
\hline

\noalign{\smallskip} 
O$^0$/H$^+$$\times$10$^{5}$		&5.2$\pm$1.1	  & 0.06$\pm$0.01 & 0.07$\pm$0.01&-		&-		&-		\\
O$^+$/H$^+$$\times$10$^{5}$		&4.3$\pm$1.6	  & 0.13$\pm$0.30 & 0.27$\pm$0.30&-		&0.4$\pm$0.9	&0.1$\pm$0.2		\\
O$^{++}$/H$^+$$\times$10$^{5}$		&3.2$\pm$0.3	  & 9.4$\pm$3.6	  & 4.8$\pm$3.5	 &3.4$\pm$1.9	&17.3$\pm$4.	&15.9$\pm$4.		\\
ICF(O)					&1.7	  	  & 1.0	  	  & 1.0	 	 &1.0		&1.0		&1.0		\\
\noalign{\smallskip} 
O/H$\times$10$^{5}$			&13.0$\pm$3.7	  & 9.6$\pm$3.7	  & 5.2$\pm$3.9	 &3.5$\pm$2.4	&17.8$\pm$5.	&16.4$\pm$5.		\\
\hline

\noalign{\smallskip} 
Ne$^{++}$/H$^+$$\times$10$^{6}$		&9.2$\pm$3.1	  &150.$\pm$80.	  &55.$\pm$30.	 &80.$\pm$60.	&30.$\pm$10.	&45.$\pm$20.		\\
ICF(Ne)					&4.1	  	  &1.0		  &1.1		 &1.0		&1.0		&1.0		\\
\noalign{\smallskip} 
Ne/H$\times$10$^{5}$			&0.38$\pm$0.20	  &1.6$\pm$1.2	  &0.6$\pm$0.3	 &0.8$\pm$0.6	&3.$\pm$1.	&5.$\pm$3.		\\
\hline

\noalign{\smallskip} 
S$^+$/H$^+$$\times$10$^{7}$		&2.3$\pm$0.8	  &-		&-		&-		&-		&-		\\
S$^{++}$/H$^+$$\times$10$^{7}$		&3.9$\pm$0.8	  &-		&-		&0.8$\pm$0.8	&0.9$\pm$0.8	&2.5$\pm$1.		\\
ICF(S)					&1.1		  &-		&-		&-		&2.5		&3.5		\\
\noalign{\smallskip} 
S/H$\times$10$^{7}$			&7.0$\pm$1.9	  &-		&-		&$>$0.8	&2.$\pm$1.	&9.$\pm$3.		\\
\hline

\noalign{\smallskip} 
Ar$^{++}$/H$^+$$\times$10$^{8}$		&9.1$\pm$8.7	  &-		&-		&19.$\pm$10.	&-		&-		\\
Ar$^{+++}$/H$^+$$\times$10$^{8}$	&-		  &-		&-		&16.$\pm$10.	&-		&-		\\

ICF(Ar)					&1.9	  	  &-		&-		&1.3		&-		&-		\\
\noalign{\smallskip} 
Ar/H$\times$10$^{7}$			&1.7$\pm$1.7	  &-		&-		&5.$\pm$4.	&-		&-		\\
\hline
\end{tabular} 
\end{center} 
\label{Tab_ion_pn} 
\end{table*} 

\begin{table*} 
\caption{ Total and ionic chemical abundances, and ionization correction factors  of \sexa\ and \sexb\ \hii\ regions,  
computed with  the ICF method. } 
\begin{center} 
\renewcommand{\arraystretch}{1.4} 
\setlength\tabcolsep{5pt} 
\begin{tabular}{llllllll} 
\hline\noalign{\smallskip} 
					& HKS3 	  	&  HKS7       	& HKS9  	& HKS14 	& SHK1 		& SHK2 	      & SHK5               \\
\noalign{\smallskip} 														                     
\hline 																                     
\noalign{\smallskip} 														                     
He$^+$/H$^+$$\times$10$^{2}$		&5.0$\pm$1.0	& 7.8$\pm$1.2	&3.0$\pm$0.5	&-		&-		&12.3$\pm$2.  &3.5$\pm$1.0      \\
He$^{++}$/H$^+$$\times$10$^{2}$		&0.09$\pm$0.01	& 0.09$\pm$0.01	&0.09$\pm$0.01	&0.3$\pm$0.1	&0.09$\pm$0.01	&0.08$\pm$0.01&0.5$\pm$0.2           	\\
\noalign{\smallskip} 											                     
He/H$\times$10$^{2}$			&5.1$\pm$1.0	& 7.9$\pm$1.2	&3.1$\pm$0.5	&-		&-		&12.4$\pm$2.  &4.0$\pm$1.2    	\\
\hline																                     
																                     
\noalign{\smallskip} 														                     
N$^+$/H$^+$$\times$10$^{7}$		&2.9$\pm$0.4	& 3.3$\pm$0.5	&4.4$\pm$0.7	&2.7$\pm$0.4	&2.0$\pm$0.3	&6.0$\pm$0.1  &2.7$\pm$0.4    	\\
ICF(N)					&5.5	  	& 2.4		&2.6		&7.2		&4.7		&2.8	      &8.7            	\\
\noalign{\smallskip} 														                     
N/H$\times$10$^{7}$			&15.7$\pm$3.0	& 7.9$\pm$2.0	&12.0$\pm$4.0	&20.$\pm$7.0	&9.5$\pm$3.5	&17.$\pm$5.   &24.$\pm$8.  	\\
\hline																                     
																                     
\noalign{\smallskip}														                     
O$^0$/H$^+$$\times$10$^{5}$		&-		& -		&-		&0.2$\pm$0.1	&-		&-	      &-               	\\
O$^+$/H$^+$$\times$10$^{5}$		&0.6$\pm$0.1	& 0.7$\pm$0.1	&0.5$\pm$0.1	&0.6$\pm$0.1	&0.6$\pm$0.1	&1.0$\pm$0.1  &0.7$\pm$0.1     	\\
O$^{++}$/H$^+$$\times$10$^{5}$		&2.7$\pm$0.9	& 0.8$\pm$0.3	&0.9$\pm$0.3	&3.8$\pm$1.3	&2.3$\pm$0.8	&1.8$\pm$0.6  &4.3$\pm$1.5     	\\
ICF(O)					&1.0	   	& 1.0		&1.0		&1.0		&1.0		&1.0	      &1.0             	\\
\noalign{\smallskip} 														                     
O/H$\times$10$^{5}$			&3.3$\pm$1.0	& 1.5$\pm$0.4	&1.4$\pm$0.4	&4.5$\pm$1.5	&3.0$\pm$0.9	&2.8$\pm$0.7  &5.0$\pm$1.6     	\\

\hline 																                     
\noalign{\smallskip} 														                     
Ne$^{++}$/H$^+$$\times$10$^{6}$		&7.0$\pm$2.0	& 1.4$\pm$0.5	&2.1$\pm$0.8	&7.7$\pm$2.7	&3.5$\pm$1.3	&7.5$\pm$2.7  &8.1$\pm$2.9     	\\
ICF(Ne)					&1.2	  	& 1.7		&1.7		&1.2		&1.3		&1.6	      &1.2             	\\
\noalign{\smallskip} 								                     
Ne/H$\times$10$^{5}$			&9.0$\pm$5.0	& 2.4$\pm$1.3	&3.5$\pm$1.9	&8.9$\pm$5.0	&4.4$\pm$2.6	&1.8$\pm$0.6  &9.5$\pm$5.0     	\\
\hline																                     
																                     
\noalign{\smallskip} 													                  
S$^+$/H$^+$$\times$10$^{7}$		&1.4$\pm$0.2	 &1.5$\pm$0.2	&2.0$\pm$0.3	&1.6$\pm$0.2	&1.3$\pm$ 0.1	&3.0$\pm$0.5  &1.80$\pm$0.2       \\  
S$^{++}$/H$^+$$\times$10$^{7}$		&6.0$\pm$1.0	 &6.0$\pm$1.7	&3.2$\pm$0.9	&3.5$\pm$0.9	&4.9$\pm$ 1.2	&7.5$\pm$2.   &6.50$\pm$1.6        \\ 
ICF(S)					&1.3	  	 &1.1		&1.1		&1.4		&1.25		&1.1	      &1.5           \\ 
\noalign{\smallskip} 													                  
S/H$\times$10$^{7}$			&9.2$\pm$2.0	 &8.0$\pm$2.0	&5.7$\pm$1.3	&7.2$\pm$1.6	&7.8$\pm$ 1.9	&11.6$\pm$3.  &12.2$\pm$3.        \\ 
\hline															                  
															                  
\noalign{\smallskip} 													                  
Ar$^{++}$/H$^+$$\times$10$^{7}$		&1.0$\pm$0.3	  &0.7$\pm$0.2	&0.8$\pm$0.2	&0.8$\pm$0.2	&1.1$\pm$ 0.3	&3.2$\pm$ 0.4	&0.5$\pm$0.1        \\
ICF(Ar)					&1.9	  	  &1.9		&1.9		&1.9		&1.9		&1.9		&1.9            \\
\noalign{\smallskip} 														                     
Ar/H$\times$10$^{7}$			&1.9$\pm$0.7	  &1.2$\pm$0.5	&1.5$\pm$0.5	&1.5$\pm$0.5	&2.1$\pm$0.7	&5.9$\pm$ 2.	&0.9$\pm$0.3       \\
\hline
\end{tabular} 
\end{center} 
\label{Tab_ion_sexabhii} 
\end{table*}

\begin{table*} 
\caption{ Total chemical abundances of \sexa\ and \sexb\ PNe computed with both ICF and CLOUDY methods. 
We also report the unweigthed average of the two results. Metal abundances are expressed in 12+$\log$(X/H). } 
\begin{center} 
\renewcommand{\arraystretch}{1.4} 
\setlength\tabcolsep{5pt} 
\begin{tabular}{lllllll} 
\hline\noalign{\smallskip} 
Name  & He/H & N/H & O/H & Ne/H & S/H & Ar/H \\ 
\noalign{\smallskip} 
\hline 
\noalign{\smallskip} 
SexA-PN			 &			&		 &	       &        	&		&	\\
ICF 		         & 0.084$\pm$0.016	& 8.35$\pm$0.2	 & 8.1$\pm$0.1 & 6.6$\pm$0.2  & 5.8$\pm$0.1 & 5.2$\pm$0.4\\
CLOUDY	 	   	 & 0.095$\pm$0.02	& 8.4$\pm$0.2	 & 8.0$\pm$0.1 & 6.7$\pm$0.2   & 5.9$\pm$0.2 & 5.2$\pm$0.2\\
  
Average 	 	 & 0.09$\pm$0.02 	& 8.4$\pm$0.2	 & 8.0$\pm$0.1 & 6.7$\pm$0.2   & 5.9$\pm$0.2 & 5.2$\pm$0.3\\  
\noalign{\smallskip} 
\hline\hline
\noalign{\smallskip} 
SexB-PN1		 &			&		 &		&		&		&	\\
ICF			 & 0.099$\pm$0.015	& 6.7:		 & 8.0$\pm$0.15	& 7.2$\pm$0.3	&-		&-	\\
CLOUDY			 & 0.10$\pm$0.02	& 6.6:		 & 8.1$\pm$0.3	& 7.3$\pm$0.3	&-		&-	\\		

Average 	 	 & 0.10$\pm$0.02 	& 6.6:		 & 8.1$\pm$0.2  & 7.3$\pm$0.3   & -		& -     \\ 
\hline  
SexB-PN2		 &			&		 &		&		&		&	\\
ICF			 & 0.096$\pm$0.016 	& 6.7:		 & 7.7$\pm$0.3	& 6.8$\pm$0.2	&-		&-	\\
CLOUDY		 	 & 0.11$\pm$0.02	& 6.6:		 & 8.0$\pm$0.2	& 7.0$\pm$0.3	&-		&-	\\

Average	 		 & 0.10$\pm$0.02	& 6.6:		 & 7.8$\pm$0.2  & 6.9$\pm$0.3	& -		& -	\\ 
\hline 
SexB-PN3		 &			&		 &		&		&		&	\\
ICF		         & 0.076$\pm$0.016 	& $>$5.5$\dag$	 & 7.5$\pm$0.3  & 6.9$\pm$0.3	& $>$4.9$\dag$	& 5.7$\pm$0.2\\
CLOUDY			 & 0.09$\pm$0.02	& 6.5$\pm$0.3	 & 7.6$\pm$0.2  & 7.1$\pm$0.2	& 5.2:		& 5.5$\pm$0.2\\

Average	 	   	 & 0.08$\pm$0.02	& 6.5$\pm$0.3    & 7.5$\pm$0.2  & 7.0$\pm$0.2	& 5.2:		& 5.6$\pm$0.2	\\ 
\hline 
SexB-PN4		 &			&		 &		&		&		&	\\
ICF	  		 & 0.095$\pm$0.014	& 6.8$\pm$0.3    & 8.3$\pm$0.1 	& 7.4$\pm$0.1	& 5.3$\pm$0.2		& -	\\
CLOUDY			 & 0.09$\pm$0.02	& 6.3$\pm$0.3	 & 8.2$\pm$0.2	& 7.4$\pm$0.2	& 5.2$\pm$0.2	        & 5.1:\\

Average 	  	 & 0.09$\pm$0.02	& 6.5$\pm$0.3    & 8.2$\pm$0.2  & 7.4$\pm$0.2	& 5.2$\pm$0.2		& 5.1:	\\ 
\hline 
SexB-PN5		 &			&		 &		&		    &		&	\\
ICF	 		 & 0.13$\pm$0.02	& 7.5$\pm$0.3	 & 8.2$\pm$0.1 	& 7.6$\pm$0.2	   & 5.4$\pm$0.2	        &-	\\
CLOUDY			 & 0.13$\pm$0.02	& 7.0$\pm$0.3	 & 8.2$\pm$0.2	& 7.6$\pm$0.2	   & 5.7$\pm$0.2	&- 	\\

Average 	 	 & 0.13$\pm$0.02 	& 7.3$\pm$0.3	 & 8.2$\pm$0.1  & 7.6$\pm$0.2	   & 5.5$\pm$0.2	& -     \\  
 \noalign{\smallskip} 
\hline 
\hline
\noalign{\smallskip} 
Average \sexb             &0.10$\pm$0.02	& 6.8$\pm$0.6	 &8.0$\pm$0.3	&7.3$\pm$0.3&	5.3$\pm$0.4&-\\ 
\noalign{\smallskip} 
\hline 
\noalign{\smallskip} 
\end{tabular} 
\end{center} 
\label{Tab_c_pn} 
\end{table*} 

The modelling with CLOUDY also allowed us to estimate some stellar
parameters as the luminosity and temperature of the exciting stars, as described 
by Magrini et al. \cite{M04} .
{\bf We remind briefly how T$_{\star}$ and L$_{\star}$ were computed in the modelling with CLOUDY.
The luminosity of the central star 
was set to reproduce the observed absolute
flux of the \oiii\ $\lambda$5007  nebular line,  measured by
Magrini et al. (\cite{m02}, \cite{m03}) via aperture photometry, and 
reddening-corrected, while the BB T$_{\star}$ was set to match the
\hei/\heii\ line ratio.} 

\begin{figure}[h!] 
\includegraphics[width=9.5 truecm, angle=0]{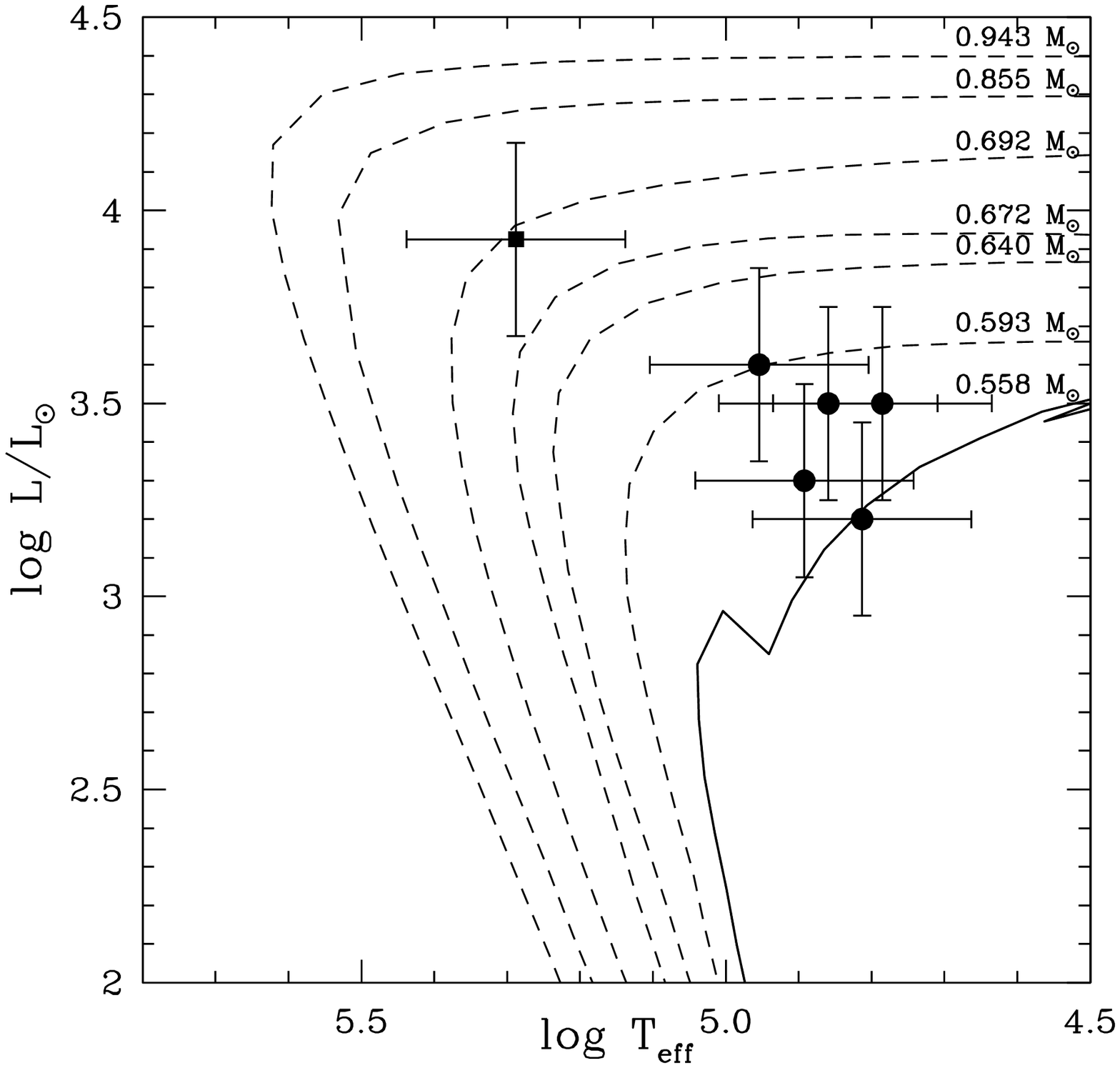} 
\caption{H-burning (dashed lines) and He-burning (continuous line)  
evolutionary tracks for metallicity Z=0.004 from Vassiliadis \& Wood
(\cite{vass}).  Final mass of the central star are quoted close to the
corresponding track.  Large symbols (circles) correspond to \sexb\ PNe
and the small one (square) indicates the \sexa\ PN. }
\label{fig_track} 
\end{figure}

Using these parameters, the PNe were located in the
$\log$T$_{\star}$-$\log$L$_{\star}$ diagram and compared with the
available H-burning and He-burning evolutionary tracks for Z=$0.004$
from Vassiliadis \& Wood (\cite{vass}), as shown in
Fig.~\ref{fig_track}.
{\bf The obtained T$_{\star}$ and $\log$(L$_{\star}$/L$_{\odot}$) are reported in 
Table~\ref{Tab_cs}. Typical errors obtained with the CLOUDY modelling are 
of the order of 0.15 for $\log$T$_{\star}$ and of 0.25 for $\log$L$_{\star}$. }  
\begin{table} 
\caption{ {\bf T$_{\star}$ and L$_{\star}$ of the PNe in \sexa\ and \sexb\ 
as derived with the modelling with CLOUDY.}}
\begin{center} 
\renewcommand{\arraystretch}{1.4} 
\setlength\tabcolsep{5pt} 
\begin{tabular}{lllc} 
\hline\noalign{\smallskip} 
Galaxy & Name   &  T$_{\star}$ (K)  & $\log$(L$_{\star}$/L$_{\odot}$)\\ 	 
\noalign{\smallskip} 
\hline\hline 
\noalign{\smallskip} 
Sextans A& PN  & 194000  &3.9 \\
\noalign{\smallskip} 
Sextans B& PN1 & 61000   &3.5  \\
& PN2 & 78000   &3.3  \\
& PN3 & 72000   &3.5  \\
& PN4 & 65000   &3.2  \\
& PN5 & 90000   &3.6  \\
\noalign{\smallskip} 
\hline 
\noalign{\smallskip} 
\end{tabular} 
\end{center} 
\label{Tab_cs} 
\end{table}

The mass of the core of the PN in \sexa\ is distinctively larger than
for the PNe in \sexb, and would imply a progenitor initial mass of
$\sim$2-3.5M$_{\odot}$, using the initial-to-final mass relation
adopted by Vassiliadis \& Wood (\cite{vass}).  The relatively large
mass of the progenitor is also typical of type~I Galactic PNe (Corradi
\& Schwarz \cite{corradi95}). 
 
The PNe of \sexb\ can be instead associated with less massive cores.
Among them, PN~5 seems to have the most massive one.  Note, however,
that the metallicity of \sexa\ and \sexb, 0.001 and 0.002,
respectively, is significantly lower than for the models of
Vassiliadis \& Wood (\cite{vass}), and therefore evolutionary tracks
for such low metallicities would be needed for a deeper analysis.
>From Fig.~\ref{fig_track} and the initial to final mass relationship
used by Vassiliadis \& Wood (\cite{vass}), we find that the central
stars of Sex~B PNe had initial masses between 0.9 and 1.5M$_{\odot}$.
Stars with these masses and with an initial metallicity Z=0.001 were
born between about 10 and 2 Gyr ago (cf. Charbonnel et
al. \cite{charbo}; \cite{charbo96}).  As the mean oxygen abundance of
PNe in \sexb\ is equal, within the errors, to that of \hii\ regions,
this would mean that no noticeable chemical enrichment has occurred
in the galaxy  since long time ago.

\subsection{Comparison with previous determinations of chemical abundances}

In Sextans~A, Skillman et al. (\cite{skillman89}) observed
spectroscopically four \hii\ regions, deriving 12 +
$\log$O/H=7.49. The same spectroscopic data were used by Richer \&
McCall (\cite{richer95}) to recalculate 12 + $\log$O/H=7.55$\pm$0.11
and by Pilyugin (\cite{pil01}) who obtained through the P-method
(Pilyugin
\cite{pilyugin00}) 12 + $\log$O/H=7.7.  An analysis of supergiant
stars by Kaufer et al. (\cite{kaufer}) derived a total metallicity
which is 1.09 dex lower than solar.  A red giants metallicity
[Fe/H]=-1.9$\pm$0.4 was derived by Grebel (\cite{grebel00}).  Dolphin
et al. (\cite{dolphin}) found through modeling of the color-magnitude
diagram a mean metallicity of [M/H]$\sim$-1.4.  These values are in
good agreement with 12 + $\log$O/H=7.6$\pm$0.2, the average oxygen
abundance derived from our observations of nine \hii\ regions,
corresponding approximately to [Fe/H]=-1.6. The transformation from
[O/H] to [Fe/H]  needed to compare our nebular oxygen abundances
with the stellar abundances was computed according to the empirical
transformation by Mateo (\cite{mateo}).   We used the solar
abundances from Grevesse \& Sauval (\cite{grevesse}) to derive [O/H]
from the measured 12 + $\log$O/H.  The oxygen abundance of the PN in
\sexa\ is significantly higher (12 + $\log$O/H=8.0$\pm$0.1), and will
be discussed in Sect.\ref{sect_pna}.

In \sexb, the observations by Skillman et al. (\cite{skillman89}) of
four \hii\ regions gave 12 + $\log$O/H=7.56, while Moles et
al. (\cite{moles90}) measured 12 + $\log$O/H=8.1 in a single \hii\
region.  Pilyugin (\cite{pil01}) recalculated both abundances
obtaining 12 + $\log$O/H=7.85 and 7.87, respectively. The average
oxygen abundance derived in this paper with eight \hii\ regions is
7.8$\pm$0.2 and from five PNe is 8.0$\pm$0.3, thus in agreement with
Moles et al. (\cite{moles90}) and Pilyugin (\cite{pil01}).

{A recent very work about spectroscopy of \hii\ regions and PNe of
\sexa\ and \sexb\ is by Kniazev et
al. (\cite{kniazev05}}
The authors derived chemical abundances of the PN and of three \hii\
regions in \sexa, and of one PN (PN3) and three \hii\ regions in
\sexb\ with the ICF method.  In \sexa, they found an oxygen abundance
of 8.02$\pm$0.05 in the PN and 7.54$\pm$0.06 in three
\hii\ regions, that, we did not observe. These \hii\ regions belong to
the same complex of HSK14, HSK16, and HSK19, for which we derived a
mean 12 + $\log$O/H=7.65$\pm$0.1.  In PN3 of \sexb\ they derived 12 +
$\log$O/H=7.47$\pm$0.16, in agreement with our value 7.5$\pm$0.2.
They also observed the same \hii\ regions for which we derived
chemical abundances with the ICF method, SHK1, SHK2, SHK5, finding 12
+ $\log$O/H 7.50$\pm$0.08, 7.55$\pm$0.06, 7.84$\pm$0.05, respectively.
These values are similar to our ICF determinations: 7.5$\pm$0.1,
7.5$\pm$0.1, 7.7$\pm$0.1.
  
\subsection{Relationships between chemical abundances}
 
\begin{figure} 
\includegraphics[width=9 truecm, angle=0]{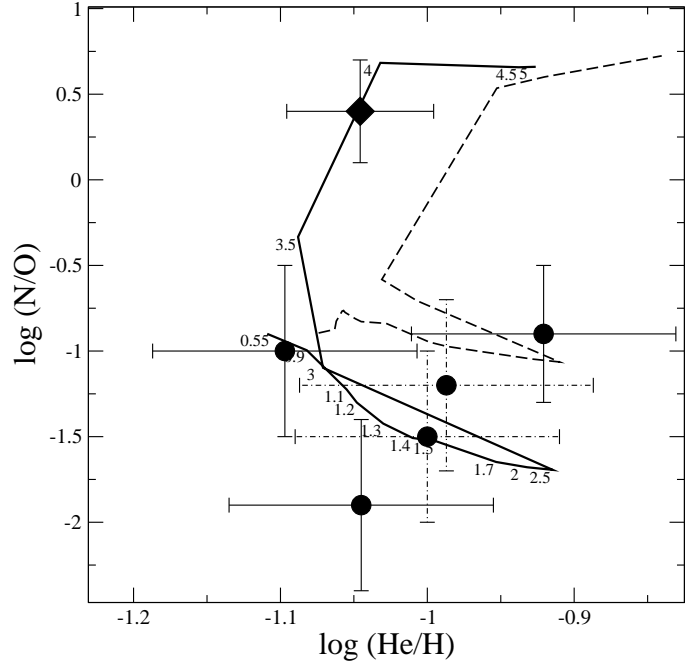} 
\caption{Relationship between He/H and N/O. The solid line is the stellar model 
for the progenitor by Marigo (2004) for Z=0.001 and the dashed line
for Z=0.004.  Values of the initial masses are indicated.  Full
circles correspond to \sexb\ PNe and the diamond is the \sexa\ PN. PNe
with uncertain N/H determination are marked with dot-dashed error
bars. }
\label{fig_he_no} 
\end{figure}

The relation between He/H and N/O is shown in Fig.~\ref{fig_he_no},
where the observed chemical abundances of PNe are compared with new
models by Marigo (2004, private communication) built for initial
metallicities Z=0.001 and Z=0.004.  Both models predict, for a
progenitor mass below 3M$_{\odot}$, an enhancement of O/H compared to
N/H, producing a N/O abundance ratio decreasing with He/H. At even
higher mass progenitors, N/O increases quickly and significantly as
the result of hot-bottom burning. The abundance of the PN of \sexa\ is
in good agreement with the model Z=0.001 and an initial mass around
4M$_{\odot}$.  Four of \sexb\ PNe match within 1-1.5$\sigma$ with the
same model, Z=0.001, and masses between 0.55 and 3 M$_{\odot}$; one of
them, PN~5, is in better agreement with the model with Z=0.004 and a
progenitor with initial mass of about 1.5M$_{\odot}$, in agreement
with the result obtained with the evolutionary tracks
(Fig.~\ref{fig_track}).  Note however that the mean metallicity of
\sexb\ is Z=0.002.

\begin{figure} 
\includegraphics[width=9 truecm, angle=0]{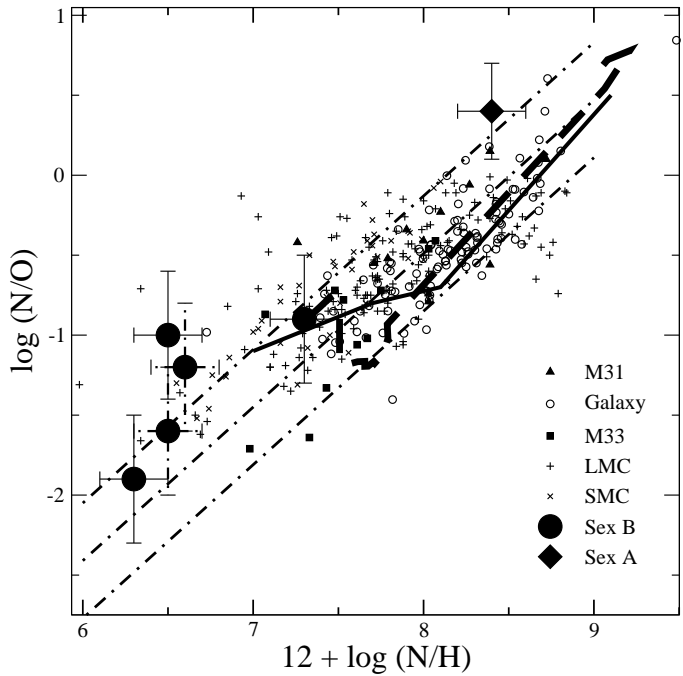} 
\caption{Relationship between N/O and N/H for PNe.  
The central dot-dashed line is the fit from Henry (\cite{henry90})
using a sample of PNe of the Galaxy, LMC, SMC and M31. The external
dot-dashed lines are placed at a distance of $\pm$3$\sigma$ from the
fit, where $\sigma$=0.1~dex.  The solid line is the model for LMC by
Groenewegen \& de Jong (\cite{groene}) built for the LMC metallicity
and several initial masses. The top-right point of the curve
corresponds to a model with an initial mass of 8.5 M$_\odot$, the
first change of slope with 2 M$_\odot$, and the last point
(bottom-left) is for 0.93 M$_\odot$.  The dashed line is the model for
Z=0.004 and initial masses from 0.8 (left) to 5 (right) M$_\odot$ by
Marigo (\cite{marigo01}).  Observed abundances are from: Galaxy
(Perinotto et al. \cite{peri04}); LMC and SMC (Leisy et al., in
prep.); M31 (Jacoby \& Ciardullo \cite{jc99}); M~33 (Magrini et
al. \cite{M04}); \sexa\ and \sexb\ (this work). PNe of \sexb\ with
uncertain N/H determination are marked with dot-dashed error bars. }
\label{fig_nh_no} 
\end{figure}

Fig.\ref{fig_nh_no} shows the relationship between N/O and N/H for PNe
in several nearby galaxies.  This relation is well defined for PNe
with a slope near to unity, suggesting that the increase of N/O with
N/H is due to the increase of N, without modifying the oxygen
abundance through the ON cycle.  Theoretical predictions of N/O vs N/H
for different masses of the progenitor stars and metallicity
(Groenewegen \& de Jong \cite{groene} for the LMC metallicity, and
Marigo \cite{marigo01} for Z=0.004) are also shown.  The theoretical
predictions of Groenewegen
\& de Jong (\cite{groene}), shown by the solid line in 
Fig.\ref{fig_nh_no}, and of Marigo (\cite{marigo01}), the dashed line,
do not extend to the low metallicities of \sexa\ and \sexb.  Because
of it, we can only say that models, in particular those of Marigo,
(\cite{marigo01}) predict the deviation from the linear relationship
of Henry (\cite{henry89}) that the observations of PNe in \sexb\ and
in the Magellanic Clouds suggest at the lowest metallicities.  From
the figure there is no indication of oxygen enrichment for PNe with
low metallicity and very low mass progenitors (M$<$1.3 M$_\odot$), as
also predicted by theoretical models of Marigo (\cite{marigo01}). The
location n the diagram of the \sexa\ PN is a clear indication of a
large nitrogen production, as expected from its higher mass
progenitor, while oxygen production is not evident from this
diagram.

\begin{figure}
\includegraphics[width=9 truecm, angle=0]{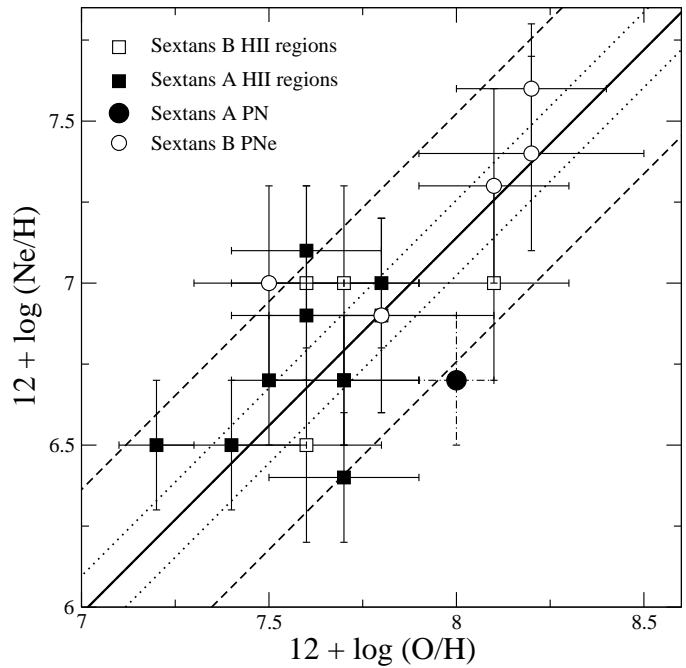}
\caption{Relationship between Ne/H and O/H. The solid  line  is the fit
built by Henry (\cite{henry89}). The dotted lines are placed at a
distance of $\pm$1$\sigma$ from the fit, and the dashed lines at
$\pm$3$\sigma$, where $\sigma$=0.1~dex as indicated by  Henry \cite{henry89}. 
Only \hii\ regions with T$_e$ measured are reported. }
\label{fig_oh_neh}
\end{figure}

Fig.~\ref{fig_oh_neh} presents the relation between O/H and Ne/H.  It
is usually believed that the progenitors of PNe do not produce
significant amount of oxygen and neon during their lifetime.  In fact,
the abundances of these two elements seem to vary in lockstep, where
the slope of the $\log$-$\log$ relation is very close to unity
(cf. Henry \cite{henry89}). This behaviour is essentially the same for
\hii\ regions and it seems independent of the host galaxy.
In Fig.~\ref{fig_oh_neh} the solid line is the fit built with a sample
of 157 PNe in the Galaxy, LMC, SMC and M~31 taken from Henry
(\cite{henry89}).  The dotted lines are placed at a distance of
$\pm$1$\sigma$ from the fit, and the dashed lines at $\pm$3$\sigma$,
where $\sigma$=0.1~dex.  The PNe and the \hii\ regions of \sexa\ and
\sexb\ appear to follow the ``universal'' relation. 

\begin{table*} 
\caption{ 
The chemical abundances of \hii\ regions in \sexa.  Where the
electron temperature is measured, values derived with both the CLOUDY
and the ICF methods are indicated. In the remaining cases, marked with
a ``$*$'', only the determinations done with CLOUDY are listed. In
computing the average abundances, upper limits of individual
determinations are not considered.}
\begin{center} 
{\footnotesize 
\renewcommand{\arraystretch}{1.4} 
\setlength\tabcolsep{5pt} 
\begin{tabular}{lllllll} 
\hline\noalign{\smallskip} 
Name   & He/H & N/H & O/H & Ne/H & S/H & Ar/H \\ 
       &	   &	 &     &      &	    &	\\	 
\noalign{\smallskip} 
\hline\hline 
\noalign{\smallskip} 
HKS2*			 &			&		 &		&		&		&	\\
CLOUDY			 &$<$0.05		& 6.1$\pm$0.3   &7.7$\pm$0.2	&6.7$\pm$0.2 	&6.1$\pm$0.2  	& 5.3$\pm$0.3\\ 
\hline  
HKS3			 &			&		&		&		&		&	\\
ICF			&0.05$\pm$0.01    	& 6.2$\pm$0.1   & 7.5$\pm$0.1   & 6.9$\pm$0.2   & 6.0$\pm$0.1   & 5.3$\pm$0.1\\
CLOUDY			&0.05$\pm$0.02    	& 6.3$\pm$0.3   & 7.6$\pm$0.2   & 6.6$\pm$0.3   & 6.1$\pm$0.2   & 5.3$\pm$0.3\\
Average  		&0.05$\pm$0.02 		& 6.2$\pm$0.2   & 7.5$\pm$0.1   & 6.7$\pm$0.2  	& 6.0$\pm$0.2  	& 5.3$\pm$0.2\\ 
\hline
HKS7			&			&		&		&		&		&	\\
ICF  			&0.08$\pm$0.01 		& 5.9$\pm$0.2 	& 7.2$\pm$0.2	& 6.4$\pm$0.2	& 5.9$\pm$0.1	& 5.1$\pm$0.3\\
CLOUDY  		&0.07$\pm$0.02   	& 6.2$\pm$0.3 	& 7.6$\pm$0.2	& 6.6$\pm$0.3 	& 6.0$\pm$0.2	& 5.4$\pm$0.3\\
Average  		&0.07$\pm$0.02   	& 6.1$\pm$0.3   & 7.4$\pm$0.2	& 6.5$\pm$0.2 	& 6.0$\pm$0.2   & 5.3$\pm$0.3\\  
\hline
HKS9			&			&		&		&		&		&	\\
ICF	    	 	&0.03$\pm$0.01  	& 6.1$\pm$0.2	& 7.2$\pm$0.2	& 6.5$\pm$0.2	& 5.8$\pm$0.1	& 5.2$\pm$0.3\\ 
CLOUDY	    		&0.03$\pm$0.02  	& 6.5$\pm$0.3   & 7.3$\pm$0.2 	& 6.5$\pm$0.3 	& 5.8$\pm$0.2	& 5.1$\pm$0.3\\ 
Average			&0.03$\pm$0.02  	& 6.3$\pm$0.2   & 7.2$\pm$0.1   & 6.5$\pm$0.2   & 5.8$\pm$0.2   & 5.1$\pm$0.3\\  
\hline
HKS14			&			&		&		&		&		&	\\
ICF  	   		& 		  	& 6.3$\pm$0.2   & 7.7$\pm$0.1	& 7.0$\pm$0.2  	& 5.9$\pm$0.2 	& 5.2$\pm$0.3\\	 
CLOUDY	  	        &$<$0.05 		& 6.3$\pm$0.3	& 7.9$\pm$0.2	& 7.0$\pm$0.3	& 6.1$\pm$0.2 	& 5.2$\pm$0.3\\ 
Average			&$<$0.05		& 6.3$\pm$0.3 	& 7.8$\pm$0.1	& 7.0$\pm$0.2	& 6.0$\pm$0.2 	& 5.2$\pm$0.3\\  
\hline
HKS16*			&			&		&		&		&		&	\\
CLOUDY			&$<$0.05		& 6.0$\pm$0.3 	& 7.7$\pm$0.2   & 6.4$\pm$0.2	& 6.0$\pm$0.2 	& 5.5$\pm$0.3\\  
\hline
HKS19*			&			&		&		&		&		&	\\
CLOUDY			& $<$0.05  		& 6.2$\pm$0.3   & 7.6$\pm$0.2   & 6.9$\pm$0.2	& 6.2$\pm$0.2   & 5.0$\pm$0.3\\ 
\hline
HKS23*			&			&		&		&		&		&	\\
CLOUDY			&$<$0.05		& 6.2$\pm$0.3 	& 7.7$\pm$0.2	& 6.7$\pm$0.2 	& 5.8$\pm$0.2   & 5.0$\pm$0.3\\  
\hline
HKS24*			&			&		&		&		&		&	\\
CLOUDY 			&$<$0.05		& 6.3$\pm$0.3   & 7.6$\pm$0.2	& 7.1$\pm$0.2 	& 6.3$\pm$0.2   & 4.7$\pm$0.3\\  
\noalign{\smallskip} 
\hline 
\noalign{\smallskip} 
Average values    &0.05$\pm$0.02&6.2$\pm$0.1&7.6$\pm$0.2&6.8$\pm$0.3&6.0$\pm$0.2&5.2$\pm$0.2\\ 
\noalign{\smallskip} 
\hline 
\noalign{\smallskip} 
\end{tabular}
} 
\end{center} 
\label{Tab_c_hii_a} 
 
\end{table*}

 \begin{table*} 
\caption{ 
The same as in Table \ref{Tab_c_hii_a}, but for \sexb.} 

\begin{center} 
{\footnotesize 
\renewcommand{\arraystretch}{1.4} 
\setlength\tabcolsep{5pt} 
\begin{tabular}{lllllll} 
\hline\noalign{\smallskip} 
Name   & He/H & N/H & O/H & Ne/H & S/H & Ar/H \\ 
       &	   &	 &     &      &	    &	\\	 
\noalign{\smallskip} 
\hline\hline 
\noalign{\smallskip} 
SHK1		     &			&		&		&		&		&	\\
ICF 	 	     &-	 		& 6.0$\pm$0.2	& 7.5$\pm$0.1	& 6.6$\pm$0.2	& 5.9$\pm$0.2	& 5.3$\pm$0.2\\ 
CLOUDY   	     &$<$0.05      	& 6.1$\pm$0.3	& 7.6$\pm$0.1	& 6.5$\pm$0.3	& 6.0$\pm$0.2	& 5.3$\pm$0.3\\ 
Average		     &$<$0.05     	& 6.0$\pm$0.2	& 7.6$\pm$0.1	& 6.5$\pm$0.2	& 5.9$\pm$0.4	& 5.3$\pm$0.3\\ 
\hline
SHK2		     &			&		&		&		&		&	\\
ICF 	  	     &0.12$\pm$0.02   	& 6.2$\pm$0.2	& 7.5$\pm$0.1	& 7.0$\pm$0.2	& 6.2$\pm$0.2	&5.8$\pm$0.2\\	 
CLOUDY    	     &0.11$\pm$0.02   	& 6.4$\pm$0.3	& 7.7$\pm$0.1	& 7.0$\pm$0.3	& 6.2$\pm$0.2	&5.8$\pm$0.3\\
Average  	     &0.11$\pm$0.02  	& 6.3$\pm$0.2	& 7.6$\pm$0.1	& 7.0$\pm$0.2   & 6.2$\pm$0.2	&5.8$\pm$0.3\\ 
\hline
SHK3*		     &			&		&		&		&		&	\\
CLOUDY		     &$<$0.05    	& 6.7$\pm$0.3	& 8.0$\pm$0.2	& -		& 6.7$\pm$0.2   & -\\ 
\hline
SHK5		     &			&		&		&		&		&	\\
ICF	       	     &0.04$\pm$0.01	& 6.4$\pm$0.2	& 7.7$\pm$0.1  	& 7.0$\pm$0.2 	& 6.0$\pm$0.2	& 4.9$\pm$0.2\\	 
CLOUDY	  	     &0.04$\pm$0.02	& 6.3$\pm$0.3	& 7.7$\pm$0.1	& 7.0$\pm$0.3	& 6.1$\pm$0.2	& 5.0$\pm$0.3\\ 
Average  	     &0.04$\pm$0.02	& 6.3$\pm$0.2	& 7.7$\pm$0.1	& 7.0$\pm$0.2	& 6.0$\pm$0.2 	& 5.0$\pm$0.3\\
\hline
SHK7*		     &			&		&		&		&		&	\\
CLOUDY  	     &$<$0.05    	& 6.4$\pm$0.3	& 7.7$\pm$0.2	& -		& 6.1$\pm$0.2	& 5.5$\pm$0.3\\
\hline
SHK6*		     &			&		&		&		&		&	\\
CLOUDY  	     &$<$0.05		& 6.6$\pm$0.3	& 7.8$\pm$0.2	& -		& 6.6$\pm$0.2	& -\\   
\hline
SHK10*		     &			&		&		&		&		&	\\
CLOUDY		     &$<$0.05		& 6.4$\pm$0.3	& 8.1$\pm$0.2	&7.0$\pm$0.2 	& 6.0$\pm$0.2 	& 5.5$\pm$0.3\\  
\hline
\hii H*		     &			&		&		&		&		&	\\
CLOUDY		     &0.05$\pm$0.02	& 6.1$\pm$0.3	& 7.8$\pm$0.2	& 6.9$\pm$0.3	& 6.0$\pm$0.2 	& 5.0$\pm$0.3\\
\noalign{\smallskip} 
\hline 
\noalign{\smallskip}
Average values       &0.07$\pm$0.04 &6.4$\pm$0.2 &7.8$\pm$0.2  &7.0$\pm$0.2 &   6.1$\pm$0.5 &  5.5$\pm$0.4\\ 
\noalign{\smallskip} 
\hline 
\noalign{\smallskip} 
\end{tabular}
} 
\end{center} 
\label{Tab_c_hii_b} 
 
\end{table*} 
 
\subsection{The distribution of chemical abundances within the galaxy}
\label{sect_sfh} 
  
As star formation in dwarf irregular galaxies is thought to be
episodic, and the solid-body nature of the rotation inhibits efficient
mixing, it is plausible that there are large abundance variations in
different regions of a galaxy (Skillman et al. \cite{skillman89}).
However, in several dwarf and low-mass galaxies it has been found that
the O/H abundance appears surprisingly homogeneous.  There are several
explanations which can be considered, as described, for example, by
Kobulnicky (\cite{kobul}).  This is true also in the case of \sexa\
and \sexb.  The abundance of their best measured element, oxygen, has a
low scatter in \hii\ regions: the average value is 7.6$\pm$0.2 for
\sexa\ and 7.8$\pm$0.2 for \sexb,  being, therefore,
consistent within errors with an uniform distribution.

Another aspect to be considered is the so called "abundance gap",
which is the difference between the abundances of \hii\ regions and
PNe (cf. Richer et al. \cite{richer97}).  Generally \hii\ regions have
larger O/H than PNe, because they belong to a younger population than
the PN progenitors, which are not considered to produce large amount
of oxygen, unless under ``special'' circumstances (see
Sect.~\ref{sect_pna}).  From the models of Richer et
al. (\cite{richer97}), this gap is expected to be as large as 0.2 or
0.3 dex, but it is a function of the metallicity and of the chemical
history of the galaxy.  Both in \sexa\ and \sexb\ this gap does not
seem to be present (for \sexa, we considered Ne/H instead of the
oxygen abundance, which might be enhanced during the evolution of the
progenitor of the PN, see discussion in Sect.~\ref{sect_pna}).  For
\sexa, its PN has a young, relatively massive progenitor and thus it
is not surprising that it has abundances similar to those of \hii\
regions.

For \sexb, the old ages of the PNe progenitors estimated in
sect.~3.3, together with their homogeneous abundances, would indicate a
very low chemical enrichment of the galaxy during a long period of
time till the present. Determining with more accuracy the length of
this period of low chemical enrichment (5-10 Gyrs?) is prevented by
the significant uncertainties and approximations done in deriving the
ages of the PN progenitors.

The average O/H abundances of HII regions in \sexa\ and \sexb\ are
slightly different (7.6$\pm$0.2 vs.  and 7.8$\pm$0.2, respectively),
but they are within the errors and computed dispersion, and therefore
no clear indication of a different chemical history can be drawn.
In fact, as \sexa\ and \sexb\ have approximately the same V-luminosity
and the same morphological type, considering the well known linear
relationship between the metallicity and the luminosity of galaxies
(see Pilyugin \cite{pil01}), they are expected to have the same O/H,
within the dispersion of the empirical relationship.  So the similar
O/H abundances, in spite of quite different SFHs, seem to confirm that
galaxies with a similar luminosity have similar metallicity and that
the luminosity-metallicity relationship represents the ability of a
given galaxy to keep the products of its own evolution rather than its
ability to produce metals (Larson \cite{larson74}).

\section{The peculiar PN of \sexa } 
\label{sect_pna} 
 
The PN of \sexa\ shows significant N/H and O/H overabundances with
respect to the \hii\ regions.  N/H is generally enhanced in PNe by
nucleosynthesis processes in the progenitor star.  The large amount of
N/H in this PN can be explained, as for Type I PN in the Galaxy
(Corradi \& Schwarz \cite{corradi95}), by a relatively high mass
progenitor which produces and dredges-up nitrogen.  On the other
hand, the abundances of neon, sulphur, and argon, which are not
expected to be varied by nucleosynthesis, are similar, within errors,
to the average values of \hii\ regions (or slightly lower,
consistently with the fact the PN belongs to a slightly older stellar
population).  

The overabundance of oxygen is more difficult to understand.  As noted
in Sect.~\ref{sect_sfh}, the O/H abundance of the \hii\ regions
studied in \sexa\ has a low dispersion, indicating that there is today
an homogeneous distribution of elements in the galaxy. These data,
however, do not allow us to exclude the existence of a metallicity
gradient through \sexa, because the \hii\ regions are located roughly
in a ring, and there are no data for the central region where the PN
is located (it should also be remembered that the PN progenitor may
have been migrated from its original place of birth).  Existing
information from the literature appears contradictory, as while the
three A-type supergiant stars located over a length of 0.8 kpc studied
by Kaufer et al. (\cite{kaufer}) have uniform Fe/H abundances, other
stars show a larger r.m.s scatter in metallicity at all ages (up to
0.4~dex for intermediate mass stars, Dolphin et al. \cite{dolphin}, or
for red giant branch stars, Grebel et al. \cite{grebel}).  However,
the hypothesis that the progenitor of the PN has formed in a region of
\sexa\ with locally enhanced metallicity seems to be ruled out (or at
least weakened) by the fact that only O/H is significantly higher
than in HII regions, while the abundance of other elements which do
not vary during stellar evolution of the PN progenitor, like sulphur,
argon, and neon, are all very similar to those of the \hii\ regions.
 
 An alternative, interesting explanation is that O/H surface enrichment
occurred during the evolution of the PN progenitor.   This was first
suggested by Leisy et al. (\cite{leisy02}) for several PNe in the Magellanic
Clouds and interpreted as a consequence of a dredge-up that
depends on the metallicity of the PN progenitor and  enriches of
oxygen its surface.  Indeed, contrary to ``classical'' stellar evolution
theories,  recent semi-analytical models  do predict  O
enrichment by core processed material brought to the surface
by the third dredge-up (cf. Marigo \cite{marigo01}, Herwig
\cite{herwig04}).  In the model of Marigo (\cite{marigo01}), the O
production depends on several parameters: the metallicity (increasing
at low metallicities), the mixing length,  and the number of dredge-up
episodes (which also increases at low metallicities because of the
larger duration of the AGB phase). At a given progenitor mass of
2M$_\odot$, Marigo  (\cite{marigo01}) found an increase of the total oxygen stellar yield
of a factor 10, when lowering the initial metallicity from solar to
Z=0.004. The oxygen enrichment becomes important only for progenitors
with M$>$1.3M$_\odot$.  In a recent paper, Herwig (\cite{herwig04})
computed intermediate mass (from 2 to 6 M$_\odot$) stellar evolution
tracks from the main sequence to the tip of the AGB for starts with
extremely low metallicities, namely Z=0.0001.  These models predict
that oxygen is enhanced in the envelopes of low metallicity AGB stars
by the third dredge-up.  The new model of Marigo (2004, private
communication), built for an initial metallicity Z=0.001 (that of
\sexa), also predicts that the third dredge-up produces some oxygen
enrichment, as seen by the slight decreasing trend of N/O with
increasing He/H in Fig.\ref{fig_he_no}.  In particular, for a PN with
a progenitor of about 3.5 M$_\odot$ and an initial metallicity of
Z=0.001, Marigo (2004) found that hot bottom burning and third
dredge-up are present.  The production of oxygen is essentially due to
the third dredge-up and it amounts to about 0.4~dex, in (surprisingly
good) agreement with the difference found between the PN and the \hii\
regions of \sexa.  Concluding, our observations of the PN in \sexa\
support the possibility that a significant oxygen surface enrichment
can occur during the evolution of highest mass PN progenitors born in
low metallicity environments.  This potentially important result is
nevertheless based on the only known example of a PN with a high mass
progenitor belonging to such a low metallicity galaxy as \sexa.
Confirming this results is an important objective of future
observational research in the field.

\section{Summary} 
\label{sect_summ} 
 
In this paper we presented an homogeneous spectroscopic study of
ionized nebulae (\hii\ regions and PNe) representative of
populations of different ages, in the two low-metallicity,
dwarf irregular galaxies \sexa\ and \sexb.
Our main results are:
\begin{itemize} 
\item[] {\sl i)} 
we have computed, via ICF and CLOUDY modeling, helium, nitrogen,
oxygen, neon, argon, and sulphur abundances both in \hii\ regions and
PNe. In particularly, the element with the most accurate
determination, oxygen, has the same average value, within the errors,
in the PNe and \hii\ regions of \sexb: 12 + $\log$(O/H)=8.0$\pm$0.3
and 7.8$\pm$0.2, respectively, indicating a very low metal
enrichment during a significant fraction of the galaxy lifetime. For
the \hii\ regions of \sexa\ we obtained 12 + $\log$(O/H)=7.6$\pm$0.2,
whereas it is 0.4~dex larger for the only PN known there;
\item[] {\sl ii)}
the dispersion of O/H in \sexa\ and \sexb\ \hii\ regions has a
r.m.s. scatter of 0.2 dex, which is consistent, within the errors,
with an uniform distribution;
\item[] {\sl iii)} 
the oxygen overabundance of the PN in \sexa\ with respect to HII
regions might indicate an efficient third dredge-up for massive,
low-metallicity PN progenitors as predicted by models of Marigo
(\cite{marigo01}, 2004).

\end{itemize}

\begin{acknowledgements}     
We are grateful to P. Marigo for the help given in the interpretation
of our results using the preliminary results of her new AGB models for
low metallicity stars.  We thank an anonymous referee for comments and
suggestions which have improved the paper.
\end{acknowledgements}


\begin{thebibliography}{} 
\bibitem[1932]{ambar} 
Ambartsumian, V. A., 1932, 
Poulkovo Obs. Circ 4, 8 
\bibitem[1992]{aparicio} 
Aparicio, A. \& Rodr\'{\i}guez-Ulloa, J. A., 1992,  
A\&A, 260, 77 
\bibitem[1996]{charbo96} 
Charbonnel, C., Meynet, G., Maeder, A. \&, Schaerer, D., 1996, 
A\&AS, 115, 339  
\bibitem[1993]{charbo93} 
Charbonnel, C., Meynet, G., Maeder, A., Schaller, G., \& Schaerer, D., 1993,  
A\&AS, 101, 415  
\bibitem[1999]{charbo} 
Charbonnel, C., D\aa ppen, W., Schaerer, D.,  Bernasconi, P. A.,  Maeder, A.,  Meynet, G.,  Mowlavi, N., 1999,  
A\&AS, 135, 405 
\bibitem[2004]{corradi} 
Corradi R. L. M. \& Magrini L., 2004,  
to appear in the proceedings of the ESO-workshop on ``Planetary Nebulae beyond the Milky Way'' 
\bibitem[1995]{corradi95} 
Corradi, R. L. M. \& Schwarz, H., 1995,  
A\&A, 293, 871 
\bibitem[1997]{corradi97} 
Corradi, R. L. M., Perinotto, M., Schwarz, H. E., Claeskens, J.-F, 1997,  
A\&A, 322, 975 
\bibitem[2003]{dolphin} 
Dolphin, A. E., Saha, A.. Skillman, E. D., Dohm-Palmer, R. C., Tolstoy, E., Cole, A. A.,  
Gallagher, J. S., Hoessel, J. G., Mateo, Mario, 2003,  
AJ, 126, 187 
\bibitem[1996]{dopita} 
Dopita, M. A., 1996,  
Proceedings of the Second Recontres du Vietnam,  
Editions Frontieres, edited by J. Tran Thanh Van, L. M. Celnikier, Hua Chon Trung, S. Vauclair, p. 295. 
\bibitem[2004]{exter} 
Exter, K. M., Barlow, M. J., \& Walton, N. A., 2004,  
MNRAS, 349, 1291 
\bibitem[1998]{ferland} 
Ferland, G. J., Korista, K. T., Verner, D. A., 
Ferguson, J. W., Kingdon, J. B., Verner, E. M., 1998, 
PASP, 110, 761 
\bibitem[1982]{filippenko82}
Filippenko, A. V., 1982, 
PASP, 94, 715
\bibitem[2000]{grebel00} 
Grebel, E. K.,  2000, 
in Star Formation from the Small to the Large Scale, ed. F. Favata, A. A. Kaas, \& A. Wilson (SP-445) (Noordwijk: ESA), 87\bibitem[2003]{grebel} 
Grebel, E. K., Gallagher, J. S. III, \& Harbeck, D., 2003, 
AJ, 125, 1926  
\bibitem[1998]{grevesse}
Grevesse, N. \& Sauval, A. J., 1998, 
SSRv, 85, 161
\bibitem[1994]{groene} 
Groenewegen. M. A. T. \& de Jong, T., 1994,  
A\&A, 282, 127 
\bibitem[1988]{gurzadyan} 
Gurzadyan, G. A., 1988, 
ApSS, 149, 343 
\bibitem[1989]{henry89} 
Henry, R. B. C., 1989,  
MNRAS, 241, 453 
\bibitem[1990]{henry90} 
Henry, R. B. C., 1990,  
ApJ, 356, 229 
\bibitem[2000]{herwig00} 
Herwig, F., 2000,  
ApJ, 605, 425 
\bibitem[2004]{herwig04} 
Herwig, F., 2004, 
ApJS, 155, 651  
\bibitem[1974]{hodge74} 
Hodge, P. W., 1974,  
ApJS, 27, 113 
\bibitem[1994]{hodge94} 
Hodge, P., Kennicutt, R. C., \& Strobel N., 1994,  
PASP, 106, 765 
\bibitem[2003]{hucht} 
Huchtmeier, W. K., Karachentsev, I. D., Karachentseva, V. E., 2003,  
A\&A, 401, 483 
\bibitem[1993]{hunter93} 
Hunter, D. A., Hawley, W. N., \& Gallagher, J. S., 1993,  
AJ, 106, 1797 
\bibitem[2004]{kaufer} 
Kaufer, A., Venn, K. A., Tolstoy, E. Pinte, C., Kudritzki, R.-P., 2004, 
AJ, 127, 2723 
\bibitem[1994]{kb94} 
Kingsburgh, R. L. \&  Barlow, M. J., 1994,  
MNRAS, 271, 257 
\bibitem[2005]{kniazev05} 
Kniazev, A. Y., Grebel, E. K., Pustilnik, S. A., 
Pramskij, A. G., Zucker, D. B., 2005, astro-ph/0502562 
\bibitem[1998]{kobul} 
Kobulnicky, H. A., 1998, 
ASPC, 147, 108
\bibitem[1981]{jacoby81} 
Jacoby, G. H. \& Lesser, M. P., 1981, 
AJ, 86, 185 
\bibitem[1999]{jc99} 
Jacoby, G. H., Ciardullo, R,, 1999,  
ApJ, 515, 169 
\bibitem[1974]{larson74}
Larson, R. B., 1974, 
MNRAS, 169, 229
\bibitem[2002]{leisy02} 
Leisy, P., Francois, P., \& Dennefeld, M., 2002,  
RMAA, 12, 140 
\bibitem[2002]{m02} 
Magrini, L., Corradi, R. L. M., Walton, N. A., et al., 2002,  
A\&A, 386, 869 
\bibitem[2003]{m03} 
Magrini, L., Corradi, R. L. M., Greimel, R., et al., 2003,  
A\&A, 407, 51 
\bibitem[2004]{M04} 
Magrini, L., Perinotto, M., Mampaso, A., Corradi, R. L. M., 2004, 
A\&A, 426, 779 
\bibitem[2001]{marigo01} 
Marigo, P., 2001, 
A\&A, 370, 194 
\bibitem[1996]{marigo96} 
Marigo, P., Bressan, A., \& Chiosi, C., 1996,  
A\&A, 313, 545 
\bibitem[1998]{mateo} 
Mateo, M. L., 1998,   
ARA\&A, 36, 435  
\bibitem[1990]{mathis}  
Mathis J.S. 1990,   
ARA\&A 28, 37  
\bibitem[1996]{mckenna} 
McKenna, F. C., Keenan, F. P., Kaler, J. B., Wickstead, A. W., Bell, K. L., \& Aggarwal, K. M., 1996,  
PASP, 108, 610 
\bibitem[1990]{moles90} 
Moles, M., Aparicio, A., \& Masegosa, J., 1990,  
A\&A, 228, 310 
\bibitem[1998]{pagel} 
Pagel, B. E. J., 1998,  
in ``Nucleosynthesis and chemical evolution of galaxies'', Cambridge Univ. Press, 1998, pag.289 
\bibitem[1998]{perinotto98} 
Perinotto, M.,  Corradi, R. L. M., 1998,  
A\&A, 332, 721 
\bibitem[2004a]{peri04} 
Perinotto, M., Morbidelli, L., Scatarzi, A.,  
2004A, MNRAS, 349, 793 
\bibitem[2004b]{P04} 
Perinotto, M., Magrini, L., Mampaso, A., Corradi, R. L. M., 2004B, 
to appear in the Proceedings of the ESO-workshop ``Planetary Nebulae beyond the Milky Way'' 
\bibitem[2000]{pil00} 
Pilyugin, L. S., 2000,  
A\&A, 362, 325 
\bibitem[2000]{pilyugin00}  
Pilyugin, L. S. 2000, 
A\&A, 362, 325 
\bibitem[2001]{pil01} 
Pilyugin, L. S., 2001,  
A\&A, 374, 412 
\bibitem[1995]{richer95} 
Richer, M. P. \& McCall, M. L., 1995,  
ApJ, 445, 642
\bibitem[2004]{richer04} 
Richer, M. P. \& McCall, M. L., 2004,  
to appear in the Proceedings of the ESO-workshop ``Planetary Nebulae beyond the Milky Way'' 
\bibitem[1997]{richer97} 
Richer, M. G., McCall, M. L., Arimoto, N., 1997,  
A\&AS, 122, 215 
\bibitem[1989]{skillman89} 
Skillman, E. D., Kennicutt, R. C., \& Hodge, P. W., 1989,  
ApJ, 347, 875 
\bibitem[1989]{stanghe89} 
Stanghellini, L. \& Kaler, J. B., 1989,  
ApJ, 343, 811 
\bibitem[2002]{stasinska02} 
Stasinska, G., 2002,  
in the Proceedings of ``Ionized Gaseous Nebulae'',  
Eds. W. J. Henney, J. Franco, M. Martos, \& M. Pe\~{n}a,  
RMAA 12, 62 
\bibitem[1991]{strobel} 
Strobel, N. V., Hodge, P., \& Kennicutt R. C. Jr., 1991,  
ApJ, 383, 148  
\bibitem[1991]{tosi} 
Tosi, M., Greggio, L., Marconi, G., \& Forcardi, P., 1991,  
AJ, 102, 951  
\bibitem[2000]{bergh} 
van den Bergh, S., 2000 in {\em The Galaxies of the Local     
Group}, Cambridge University Press, vdB00    
\bibitem[1994]{vass} 
Vassiliadis, E. \& Wood, P. R., 1994, 
ApJS, 92, 125 

 
\end{thebibliography}
\end{document}